\pdfoutput=1

\PassOptionsToPackage{hyphens}{url}

\documentclass[acmsmall, screen=true, nonacm]{acmart}
\def\BibTeX{{\rm B\kern-.05em{\sc i\kern-.025em b}\kern-.08emT\kern-.1667em\lower.7ex\hbox{E}\kern-.125emX}}
\acmJournal{TOPS}
\usepackage{tabularx}
\usepackage{subfigure}
\usepackage[inline]{enumitem}
\usepackage{pifont}
\usepackage{attachfile2}

\usepackage{pgfplots}

\usepackage[colorinlistoftodos,prependcaption,textsize=tiny,disable]{todonotes}

\newcommand{\todoI}[1]{\todo[inline,size=\normalsize]{#1}}

\begin{document}
\title{An Architecture for Distributed Digital Identities in the Physical World}

\author{René Mayrhofer}
\email{rm@ins.jku.at}
\orcid{0000-0003-1566-4646}

\author{Michael Roland}
\email{michael.roland@ins.jku.at}
\orcid{0000-0003-4675-0539}

\author{Tobias Höller}
\email{tobias.hoeller@ins.jku.at}
\orcid{0000-0002-0132-6817}

\author{Philipp Hofer}
\email{philipp.hofer@ins.jku.at}
\orcid{0000-0002-7705-9938}

\author{Mario Lins}
\email{mario.lins@ins.jku.at}
\orcid{0000-0003-1713-3347}

\affiliation{%
	\institution{Johannes Kepler University, %
    Christian Doppler Laboratory for Private Digital Authentication in the Physical World}
	\city{Linz}
	\country{Austria}
}

\begin{abstract}
Digital identities are increasingly important for mediating not only digital but also physical service transactions. 
Managing such identities through centralized providers can cause both availability and privacy concerns: 
single points of failure and control are ideal targets for global attacks on technical, organizational, or legal fronts. 
We design, analyze, and build a distributed digital identity architecture for physical world transactions in common scenarios like unlocking doors, public transport, or crossing country borders.
This architecture combines (biometric and other) sensors, (established and upcoming) identity authorities, attribute verifiers, and a new core component we call the \emph{Personal Identity Agent (PIA)} that represents individuals with their identity attributes in the digital domain. 
All transactions are conducted in a completely decentralized manner, and the components for which we currently assume central coordination are optional and only used for assisting with service discovery and latency reduction. 
We present a first protocol between these parties and formally verify that it achieves relevant security properties based on a realistic threat model including strong global adversaries.
A proof-of-concept implementation demonstrates practical feasibility of both architecture and initial protocol for applications that can tolerate end-to-end latencies in the range of a few seconds.
\end{abstract}



\maketitle

\section{Introduction}
\label{sec:introduction}
Digital identities have become indispensable and are an essential cornerstone of many services, applications, and social or economical use cases in our daily life. 
A digital identity is the (partial) digital representation of a subject, like a human individual; it opens new opportunities when bridging the gap between the physical and the digital world, but also poses new challenges. 
Common examples of digital identities already used by many individuals are digital certificates, chip cards including biometric information, digital driving licenses, or tickets for public transportation stored on a user's smartphone. 
Most of these readily available digital identities are stored in a controlled environment like a smart card chip or the smartphone carried by a user. 
Therefore, digital identities still require their owners to carry physical tokens with them for successful authentication. 
Wouldn't it be more convenient and natural to just simply get on the public transport vehicle because you already have permission to use it, without carrying further identification documents or hardware?

This simplified scenario requires a system that recognizes an individual, performs authorization based on biometric information, and verifies additional attributes (i.e., permissions) to grant access to a service. 
An example for such a system that has already been implemented is Aadhaar~\cite{bib:aadhaar} (Unique Identification Authority of India). 
This system stores an extensive set of biometric reference data of most citizens in a centralized database. 
Public and even private services can query this biometric information to authenticate an individual in order to derive further attributes like the authorization to use a specific service. 
However, we believe that such systems pose critical risks, especially when considering privacy concerns.
A single vulnerability in centralized and highly sensitive database systems can lead to fatal consequences as shown by Khaira~\cite{bib:2018:khaira:aadhaar-breach} where attackers gain unrestricted access to more than 1 billion Aadhaar entries. 
Additionally, operators of biometric services can easily identify all Aadhaar participants without their consent and can prevent their use in a form of personalized censorship of identities.

This paper presents \emph{Digidow}\footnote{\url{https://www.digidow.eu/}}, a novel approach towards a digital identity system for the physical world with decentralized, user-centered design and strong privacy guarantees. 
This is explicitly in contrast to: the multitude of digital ID systems for digital transactions (such as the omnipresent web login with its many approaches and standards); centralized ID systems (for digital and/or physical interactions); and (potentially decentralized) digital ID systems relying on users carrying electronic devices (such as smartphones). All these are well-established subsets of dealing with identity information and are orthogonal to the scope of this article.

In particular, we note that we assume online connectivity between the main components of our proposed architecture: sensors, verifiers, and Personal Identity Agents (PIAs) and at least eventual connectivity to auxiliary components like issuing authorities and registries. 
This online setting is required to alleviate the need for users carrying documents or devices.
It is possible to support offline scenarios without live Internet access through, e.g., regularly synchronizing with a PIA instance running on the user's smartphone and local network communication with the respective local verifier and sensor.
However, this re-introduces the assumption of mobile devices and is therefore outside the focus of this work.

\paragraph{Contributions}
\begin{itemize}
	\item We for the first time introduce a digital identity system architecture (\autoref{sec:architecture}) that offers both decentralized \emph{creation}\footnote{Creation, also called provisioning, of digital identity attributes is the focus of many Self-Sovereign-Identity (SSI) approaches.} and \emph{use} of digital credentials \emph{stored on cloud instances} (rather than mobile devices) with \emph{biometric authentication} of \emph{real-world interactions} and a \emph{strong focus on privacy}. To the best of our knowledge, this combination of aspects has not been provided by any previous digital ID system.
	\item We describe a first, abstract \emph{protocol for decentralized presentation of cloud-hosted digital ID} (\autoref{sec:abstract-protocol-definition}) with consistent communication over the Tor network overlay to provide network-level privacy. Creation and provisioning of digital IDs is envisaged to rely on existing (typically nation state backed) ID infrastructure as trust anchors and is therefore out of scope of this article.
	\item We provide an in-depth analysis of the most challenging aspects of privacy-by-architecture digital identity systems and how they can be addressed with existing technologies (\autoref{sec:challenges}).
	\item \emph{Formal verification} of our abstract protocol with \textsc{Tamarin} in context of a practical threat model and assumptions based on these existing technologies proves that it achieves the relevant security goals (\autoref{sec:formalverification}); verification of privacy goals is subject to future work.
	\item A \emph{proof-of-concept implementation} in Rust demonstrates practical feasibility of both the architecture and our initial protocol version for many potential applications, particularly for physical access (\autoref{sec:prototype}). However, scenarios with low latency requirements such as high-rate public transport gate control cannot yet be achieved with current technologies in our decentralized architecture when network privacy against strong global adversaries is considered as a requirement.
\end{itemize}

\pagebreak

\section{Digidow Architecture}
\label{sec:architecture}
This section introduces relevant stakeholders, the underlying threat model that serves as the foundation for essential design decisions, and the overall architecture of \emph{Digidow}. 

\paragraph*{Note on terminology:} Within the scope of this paper, we use the term \emph{authentication} to primarily refer to ascertaining the identity of a stakeholder, both in the sense of individuals and their physical presence on the one hand and (mostly cryptographic) proof of communications endpoint properties for digital implementations. 
We use \emph{authorization} to describe the higher-level decision-making process if a party should be granted access to an interaction; this will generally include authentication as a necessary step to ascertain which party this authorization is referring to. 
However, as related work may occasionally use these terms with different nuance---particularly by applying the term authentication to include authorization aspects---especially \autoref{sec:comparison} should be read with some flexibility on these terms.

\subsection{Stakeholders}
\begin{enumerate}[start=1,label={S\arabic*}]
\item\label{stakeholder:owner} \textbf{Identity Owner:} An identity owner is an \emph{individual} who wants to use a digital identity in order to access digital and/or physical services. 
Their ideal digital identity system is secure, always available to them (even if they don't have any physical identity tokens with them), easy to use, and keeps them in control over their own data. 
Control in this context means that identity owners want to decide where their data is stored and who it is shared with. 
This implies following the principle of data minimization to ensure that identity owners only share the minimal set of information necessary to prove a particular claim, e.g.\ ``age over X years'' instead of their full date of birth to gain age-gated permissions\footnote{Many practical use cases like purchasing alcohol or entering bars currently involve scanning a full identity document like a driving license, sharing far more identity attributes than necessary.}.
\item\label{stakeholder:issuer} \textbf{Identity Issuer:} An identity issuer, also called issuing authority, is responsible for issuing attributes associated with a particular digital identity. 
To verify whether an individual is authorized to receive the respective attributes, the issuer needs a workflow for the verification and provisioning part. 
\item\label{stakeholder:verifier} \textbf{Identity Verifier:} An identity verifier wants to protect certain assets from unauthorized access. 
Therefore, a verifier wants to learn certain attributes about the identity owner trying to use its services in order to make a decision if they should be granted access. 
Verifiers need to make these decisions quickly in order to handle large numbers of users. Hence, learning attributes has to be a fast and efficient process. 
Additionally, the provided identity information must be reliable (i.e., tamper-proof and tied to a trusted identity issuer). 
\end{enumerate}

\subsection{Threat Model}
\label{sec:threatmodel}
The primary security objective of \emph{Digidow} is to provide a digital identity system that ensures the identity remains secure against any form of compromise or unauthorized interactions by an adversary (e.g., spoofing, unauthorized usage, or modification), while also addressing privacy concerns. 
Based on this objective, our threat model addresses key security aspects, setting the foundation for relevant design decisions in our architecture and shaping the process from digital identity issuance to privacy-focused identity usage, ensuring that only relevant information is shared with the verifier.

\subsection{Trust Assumptions}
\label{sec:threatmodel:trustassumptions}
We make the following trust assumptions for our threat model: 
\begin{enumerate}[start=1,label={A\arabic*}]
	\item All stakeholders act in their own interests, but are in general mutually untrusted. \label{trust:stakeholder-interest}
	\begin{enumerate}[label={\alph*}]
		\item\label{trust:stakeholder-interest:owner} Owners (\ref{stakeholder:owner}) are trusted to protect their keys and to not share them with other parties.
		More specifically, owners do not share access to their PIA (cf.\ section~\ref{sec:component:pia}) with other individuals.
		\item\label{trust:stakeholder-interest:pia} Owners (\ref{stakeholder:owner}) share only the minimum set of attributes required during an interaction.
		\item\label{trust:stakeholder-interest:issuer-verifer} Issuers (\ref{stakeholder:issuer}) are trusted by verifiers to vet the association of identity attributes to identity owners and to protect their signing keys against abuse.
		\item\label{trust:stakeholder-interest:issuer-owner} Issuers (\ref{stakeholder:issuer}) are trusted by owners not to create fake/spoofed/shadow credentials with their (biometric) identity data to other individuals.
		\item Verifiers (\ref{stakeholder:verifier}) are not trusted to keep any data---explicitly including identity attributes---they receive during an interaction confidential.
		\item\label{trust:stakeholder-interest:verifer} Verifiers (\ref{stakeholder:verifier}) and issuers (\ref{stakeholder:issuer}) may collude and share data among each other, including but not limited to the purpose of linking and tracking identity owner interactions. Owners do not intentionally collude with verifiers.
	\end{enumerate}
	
	\item\label{trust:secure-hardware} Secure hardware works as specified and is not broken (cf.\ section~\ref{sec:prototype:hw-rot}).
	\begin{enumerate}[label={\alph*}]
		\item Sensor, issuer, and PIA private keys are stored and processed in secure hardware and---with sufficient probability---cannot be cloned undetectedly by an adversary even after compromising the device operating system.
		\item Devices utilizing a secure hardware root-of-trust provide authentic attestation information. 
		Authenticity of such devices can be independently verified, ensuring that tampered code gets detected and sensitive information, such as biometric data, cannot be transmitted to an adversary-controlled network or device without detection. 
	\end{enumerate}
	
	\item\label{trust:cryptography} Cryptographic primitives (cf.\ section~\ref{sec:prototype:cryptography}) work as specified and are not broken. If quantum computing can be used to break current (asymmetric) primitives, we assume PQC variants will be used.
	
	\item\label{trust:tor} Tor network privacy assumptions (cf.\ section~\ref{sec:prototype:network-privacy}) can only be broken with small probability for individual users.
	Global adversaries cannot reliably track network connections at will.
	
	\item\label{trust:biometry} Biometric recognition (cf.\ section~\ref{sec:prototype:biometry}) models are sufficiently accurate and sensors can detect spoofed readings so that targeted attacks against specific identity owners only succeed with small probability.
	
	\item\label{trust:implementation} Code implementations (cf.\ section~\ref{sec:prototype:implementation}) follow secure design principles and generally work as specified. Software vulnerabilities are detected and fixed in a timely manner. 
	While we fully acknowledge that such vulnerabilities are among the most likely reasons for invalidating security and privacy assumptions, this is considered an orthogonal problem not specific to digital identity systems and therefore (mostly) out of scope of this article.
\end{enumerate}

\subsection{Threats}
\label{sec:threatmodel:threats}
\begin{enumerate}[start=1,label={T\arabic*}]
\item \textbf{Identity theft:}\label{threat:identity-theft}
	Attackers can try to link valid attributes with incorrect digital identities by either stealing and reproducing identifying information, or by tricking identity issuers into linking attributes with the wrong identifying information. 
	Both scenarios result in more than one individual being perceived as the same identity owner. 
	Centralizing storage of identity attributes---for example Aadhaar---is especially dangerous in this context because breaching a single database can compromise identifying information for all users at once~\cite{bib:2018:khaira:aadhaar-breach}. 
\item \textbf{Identity modification:}\label{threat:identity-modification}
	Attackers may modify attributes tied to a digital identity to reduce or expand identity owners capabilities. 
	This results in identity owners either not being able to access services they should be able to access or being able to access services they should not be able to access. 
\item \textbf{Data collection and aggregation:}\label{threat:data-collection}
	A state, an identity verifier, an issuing authority, or other interested parties may collect large amounts of (meta)data produced by an identity system to establish links that reveal information about individuals beyond what they intended to share~\cite{bib:2006:clauss:linkability-in-identity-management}. 
	This data could be used for personal advertisements, %
	social profiling~\cite{bib:2019:bilal:socialprofiling}, mass surveillance~\cite{bib:2022:qian:surveillance-in-china}, discrimination, or repression~\cite{bib:2013:fourcade:economic-classification,bib:2013:rice:credit-scoring-discriminatory-effects}. 
\item \textbf{Oversharing of identity data:}\label{threat:data-oversharing} 
	An identity verifier may receive more information about an identity owner than necessary because, in general, identity documents contain more identity data than needed for a specific transaction. This is relevant when the whole identity document is either shared or not instead of being able to select specific identity attributes.
\item \textbf{Denial-of-service:}\label{threat:denial-of-service} 
	An attacker may disturb components required for proper functioning of the identity system to prevent a stakeholder to perform related tasks. This threat only includes targeted attacks against individual stakeholders; disruptive attacks against the entire system are generic to all distributed systems and out of scope of this article. 
\end{enumerate}

We argue that this combination of trust assumptions and threat model closely matches related ones, including the eIDAS ARF currently being specified~\cite{bib:2025:arf-1.6.1}, although on a more abstract level.

In the following, we design a self-sovereign identity system to address stakeholder requirements and these threats under the trust assumptions listed above. 
This architecture presents a refinement of an earlier, initial proposal of expected components~\cite{bib:2020:mayrhofer:digidow-architecture-poster}.
The refined architecture is based on the outcomes of an evaluation of potential options for decentralization in privacy-preserving digital identity systems for implicit digital authentication in the physical world~\cite{bib:2023:roland:hmdw}.

\subsection{Components}
\label{sec:architecture:components}

\subsubsection{Personal Identity Agent (PIA)}
\label{sec:component:pia}
Every identity owner participating in a digital identity system needs a place to store their identity data as a set of attributes. 
Storing this information in a central database (like Aadhaar) is possible, but limits the level of control the identity owner can exert over their data due to the additional capabilities held by the database operator; an operator is always in a privileged position to eavesdrop on, modify, or deny use of data held in their service, potentially overriding end-user intentions.
Alternatively, the data could be stored on a smart card or similar secure storage device under direct physical control of the individual.
This grants the owner of such a device full control over who can access the data on it, but comes with a series of drawbacks: 
losing or forgetting their digital identity storage would lock users out of services that they should be able to access and newly issued attributes must somehow be written to the user-controlled storage. 

The approach chosen in \emph{Digidow} is to have identity owners store their identity information on a service (which may or may not be a separate physical device) under their control, but explicitly with online connectivity, such as smartphones, home routers,  or home servers.
Alternatively, identity owners might choose an external cloud service provider offering appropriate security guarantees to ensure confidentiality and integrity when handling identity information. 
To minimize reliance on trust and enhance verifiability of whether these security guarantees are given, recent CPU models enable leveraging remote attestation capabilities, such as those provided by the AMD SEV-SNP~\cite{bib:amd-sev-snp} or Intel TDX~\cite{bib:intel-tdx}. 
In \emph{Digidow}, we refer to the selected service as \emph{Personal Identity Agent (PIA)}~\cite{bib:2020:mayrhofer:digidow-architecture-poster}, to indicate that it can take a more proactive role in handling identity attributes compared to a \emph{wallet} as pure credential storage as envisaged, e.g., in the current European Digital Identity (EUDI) effort~\cite{bib:2023:arf-1.1.0}.
However, the primary intention of this component is similar in retaining an individual's control over their data while also enabling remote access to digital identities. 
Note that we do not assume always-on connectivity (being able to disable your own PIA is part of controlling your own identity), but digital identities held by a PIA are only available if that PIA is online.

Naturally, this approach also comes with a set of disadvantages, most notably the fact that this requires all data owners to operate a publicly available service holding highly sensitive information; most users have neither the skills nor inclination required to maintain the security and availability of such a service.  
Section~\ref{sec:challenges} will discuss how PIA implementations could address these issues. 

\subsubsection{Issuing Authority}
\label{sec:component:issuing-authority}
Every identity issuing authority participating in a digital identity system needs authoritative knowledge about certain attributes of potential identity holders, a method to correctly identify them, and an infrastructure capable of creating valid digital representations for such attributes. 
They are responsible to ensure that all attributes they provision to a PIA actually belong to the individual represented by the PIA.

We do not make any specific assumptions about the provenance of issuing authorities and attributes they assign. Whereas national or state government organizations are obvious candidates, public travel services providers, insurance companies, employers, store chains, etc. are other examples for potential attribute issuers. 
Our proposed architecture also allows for completely self-issued credentials with attributes defined by the individuals they describe. 
However, it is up to each verifier how much trust they assign to each issuing authority.
Self-sovereignty in our approach derives from control over the \emph{use} of identity attributes through the PIA, whereas issuer trust is completely orthogonal in practical systems.

\subsubsection{Verifier}
\label{sec:component:verifier}
Verifiers obtain reliable information about individuals by verifying attributes provided by their PIAs. 
This enables verifiers to make decisions based on the values of those attributes as well as the authority responsible for issuing them. 
A verifier could, for example, require an age attribute signed by a governmental issuing authority to confirm that they can legally sell them alcohol or restrict access to corporate facilities by unlocking doors only for authorized personnel with permissions issued by their employer. 

\subsubsection{Sensor}
\label{sec:component:sensor}
Sensors are necessary if a digital identity system enables transactions in the physical world~\cite{bib:2023:roland:hmdw}. 
They are responsible for linking an individual requesting a service in the physical world to their PIA representing them in the digital world. 
This link is established by collecting biometric features of individuals, which enables PIAs to prove to verifiers that they are owned by an individual with the same biometric features as the person currently located in front of the sensor. 
They have to be installed at every location in the physical world where users should be able to use the digital identity system. 
A reasonable assumption is that such sensors are mostly operated by verifiers requiring identity attributes from individuals at certain locations. 
Since a single biometric modality would not fit every scenario, there are no restrictions on modalities to be used by sensors.

\subsubsection{Actuator}
\label{sec:component:actuator}
Actuators are responsible for translating the decisions made by verifiers into actions in the physical world~\cite{bib:2023:roland:hmdw}. 
This could include unlocking a door or showing a green light. 
Since the focus of this work is on transactions in the physical world, the following text no further distinguishes between a verifier and its associated actuator(s).

\subsection{Data Flow}
\label{sec:architecture:data_flow}
\begin{figure}
	\includegraphics[width=\columnwidth]{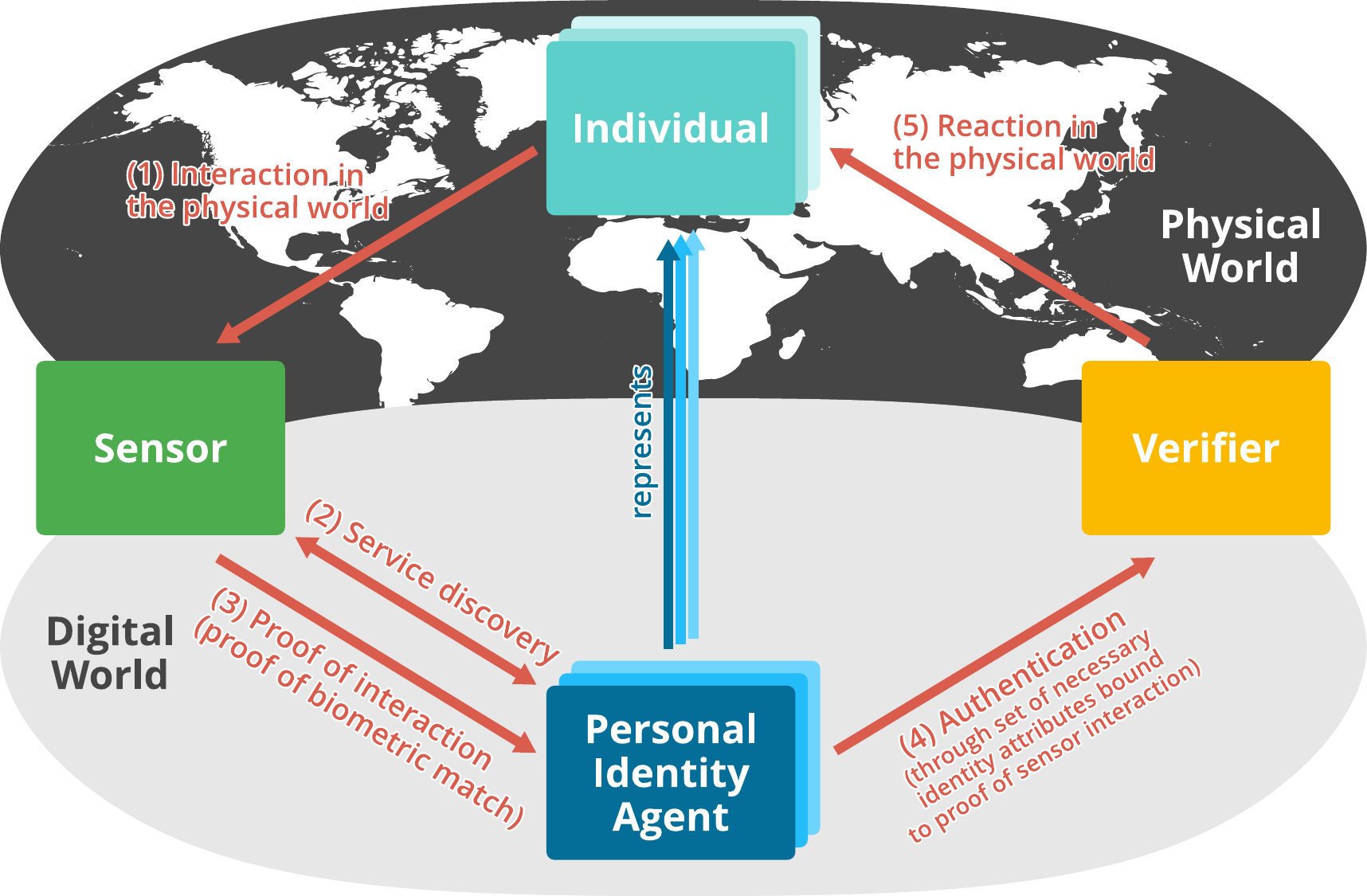}
	\caption{Steps of a Digidow transaction (refined from~\cite{bib:2023:roland:hmdw})}
	\label{fig:digidow-overview}
\end{figure}
Figure~\ref{fig:digidow-overview} visualizes the essential steps of a Digidow transaction to use a digital identity (a detailed description of protocol flows follows in section \ref{sec:abstract-protocol-definition}). 
The transaction begins with an individual interacting with a sensor. 
The sensor extracts biometric information about the individual and tries to find a PIA responsible for the individual in front of the sensor.
If a valid and trustworthy PIA is found, the sensor forwards to the PIA a signed confirmation that matching biometrics have been detected by the sensor at this point in time. 
The data owner/its PIA has the option to abort the transaction before any verifier had an opportunity to learn anything about the individual. 
This ensures that individuals using Digidow retain full control over their personal information (necessary security and privacy guarantees for biometric information collected by the sensor are discussed in more detail in section~\ref{sec:challenges:unlinkability:trusting-sensors}). 
If the PIA continues the transaction, it provides the verifier with the required attributes along with the confirmation of the sensor that its owner is currently located at a specific sensor. 
The verifier then checks if the required attributes were provided, if they have acceptable values, if they were issued by a trustworthy issuing authority, if they were issued for the biometrics detected by the sensor, and if the sensor is trusted. 
Depending on the results of those checks, the verifier triggers an appropriate actuator resulting in a reaction in the physical world, either granting or denying the individual access to a specific service.

\section{Challenges}
\label{sec:challenges}
Building a system as presented in section~\ref{sec:architecture} becomes challenging when considering requirements like privacy---which are integral to any digital identity system addressing the threats presented in section~\ref{sec:threatmodel}. 
This becomes especially difficult when functional requirements like performance, scalability, and usability are also considered. 

\subsection{Privacy via Unlinkability}
\label{sec:challenges:unlinkability}
Whereas privacy is a good requirement to strive for, it is hard to quantify the privacy properties of different implementations.
In Digidow, we want to achieve strong privacy guarantees wherever possible by requiring \emph{unlinkability} to minimize privacy risks. 
Unlinkability is defined as the inability of an observer to decide if two items of interest (IOI) within a particular set of information are related or not~\cite{bib:2010:pfitzmann:terminology-for-data-minimization,bib:2013:rfc6973:privacy-considerations}.
Both, sets of information and items of interest, vary depending on which aspect of the Digidow architecture is being analyzed but identifying and defining them is relatively simple. 
Achieving unlinkability prevents attackers, whether being passive observers or actively malicious, from obtaining more information about an individual than intentionally shared (threat \ref{threat:data-collection}) via their PIA---even if multiple such actors share IOIs and transaction details with each other (commonly referred to as \emph{collusion}). 
This ensures that the owner of a PIA retains control over their data and provides a solid foundation for an identity system protecting the privacy of its users~\cite{bib:2023:etsi-tr119476:electronic-signatures-and-infrastructures}. 

This section discusses several different aspects of unlinkability with regard to their items of interest as well as how unlinkability could be achieved in all these scenarios. 

\subsubsection{Network Privacy}
\label{sec:challenges:unlinkability:network-privacy}
The first unlinkability issue within the Digidow architecture is found within its distributed nature. 
Sensors and verifiers need to communicate with PIAs and these network requests must be considered IOIs. 
Even with perfectly working encryption, network requests are easily linkable based on timings and protocol metadata (like IP addresses or TCP ports).
Global passive adversaries observing network traffic could easily track which sensors and verifiers a PIA is communicating with~\cite{bib:2013:macaskill:nsa-files}.
Since services in the physical world are typically tied to physical locations, this results in a fairly accurate movement and behavior profile of an individual. 

Whereas it is impossible to remove linkable metadata without completely rewriting network protocols~\cite{bib:2015:chen:hornet}, there are ways to prevent attackers from linking different network requests by obscuring metadata. 
Established overlay networks like Tor~\cite{bib:2004:dingledine:tor}, I2P~\cite{bib:i2p} or Nym~\cite{bib:2021:diaz:nym-whitepaper} can achieve this, but their approach of routing traffic via multiple nodes incurs latency and reliability penalties.
This makes it more difficult to achieve the necessary reaction times needed for individuals to perceive the system as useful.  

\subsubsection{Sensor Trust}
\label{sec:challenges:unlinkability:trusting-sensors}
Sensors are in a unique position within the Digidow architecture because they have access to a highly confidential IOI: the biometric information of individuals. 
This biometric information must be linkable (comparable to other biometric data to decide if they are related or not) in order to fulfill its purpose---demanding unlinkability for this IOI would render biometry useless. 
Therefore, Digidow has to settle for the next best option:
ensuring that sensors within the Digidow architecture can be trusted to not reveal information about detected biometric information. 

To achieve this, hardware \emph{roots of trust} (RoT) such as \emph{trusted platform modules} (TPM) must be included in every sensor. 
These components enable remote attestation about both the hardware trust anchor and software run by the sensor. 
Trustworthy sensors within the Digidow architecture must be able to attest that they run trusted source code built in a reproducible way with a standardized configuration that
\begin{enumerate*}[label=(\alph*)]
\item provides timely and untampered sensor readings, ideally with anti-spoofing mitigations for the respective biometry; and 
\item neither stores information persistently on the system; 
\item nor sends any information via the network to any other party but those PIAs responsible for the human individuals currently interacting with the sensor. 
\end{enumerate*}
This also serves a second purpose as it assures PIAs and verifiers that a sensor is not maliciously reporting falsified sensor data. 

It should be acknowledged at this point that there are already cities overflowing with cameras~\cite{bib:2021:hao:surveilling-surveillance} doing face recognition for centralized surveillance systems. 
In these locations the additional leakage of biometric information by Digidow sensors would be negligible. 
As this is not the case everywhere however, in Digidow we aim to find out how the benefits provided by biometric sensors can be enjoyed while still protecting the privacy of their users. 

\subsubsection{Biometric Privacy}
\label{sec:challenges:biometric-privacy}
Since Digidow relies on biometric features for identification in the real world, sensors need something to compare their collected biometric information to. 
Whereas it is easy to compare two sets of biometric features if you know both of them, doing so between two parties without disclosing anything about their biometric templates to each other is a challenge. 
Privacy-preserving biometric schemes~\cite{bib:2016:natgunanathan:privacy-in-biometrics} striving to realize this goal are an active field of research, but they have not yet managed to achieve the necessary levels of accuracy and computational complexity needed in a digital identity system. 
Demanding unlinkability makes this challenge even more difficult and therefore not likely to be achieved in the near future. 
Conveniently, the hardware \emph{root of trust} introduced in section~\ref{sec:challenges:unlinkability:trusting-sensors} already provides PIAs with assurances that the sensor will handle biometric data confidentially. 
Assuming \ref{trust:secure-hardware}, this allows PIAs to simply forward their biometric information to sensors and trust them to do the comparison correctly. 

\subsubsection{Cryptographic Privacy}
\label{sec:challenges:cryptographic-privacy}
Cryptography provides another layer of potential linkability, with signatures and cryptographic keys as the most common IOIs. 
All messages and data exchanged between Digidow components need to make sure not to include such linkable elements. 
Whereas message parts created by the PIA are under its direct control, other parts created at other stakeholders and only forwarded by the PIA are potentially problematic for unlinkability.
We refer to section~\ref{sec:prototype:cryptography} for particular issues with signatures by issuing authorities.
Any form of identifier, even when not transmitted directly but, e.g.\ implied by the cryptographic signature scheme and therefore available to the verifier, must be avoided.
A recent consensus statement in response to the EU Identity Wallet (EUDI) Architecture Reference Framework (ARF) under development at the time of this writing~\cite{bib:2025:arf-1.6.1} details the issue of unlinkability in cryptographic signatures~\cite{bib:2024:arf-feedback} in line with how we define the term within the scope of this paper.

Another aspect of cryptographic unlinkability revolves around linking disclosed attributes. 
A verifier must be able to confirm that different attributes from different issuing authorities provided by a PIA actually refer to the same individual. 
However, a verifier should not be able to obtain more information than what was intentionally linked. 
If an issuing authority issues an age attribute as a date of birth, it could be used to verify if an individual is an adult or not, but it would also make individuals linkable. 
This could be solved by obtaining coarse-gained attributes---e.g. ``older than 18''---from the issuing authority, but this creates a new problem:
How would an issuing authority know in advance which age attributes it should issue? 
There are a few common age limits (16, 18, 21, etc.) that may be pre-issued, but it is difficult to find a suitable globally valid set of such age checks (e.g., sometimes verifiers need to check if an individual was born before a certain cut-off date).
These scenario-specific attributes cannot easily be issued in advance. 
Issuing them on the fly is not an option either, because this would enable verifiers to link their information with issuing authorities to find out who used their services. 
The only viable alternative is a solution where issuing authorities issue only a single attribute and PIAs have the capability to either selectively disclose only parts of the attribute (i.e. just the year of birth from a full date of birth) or to construct verifiable claims about their issued attributes (i.e. born before a certain date) themselves.
Range proofs or even more complex arithmetic/logical statements about attributes would be desirable in the future~\cite{bib:2005:boneh:2dnf-formulas}, but are difficult to achieve with established cryptography. 

\subsection{Network Discovery}
\label{sec:challenges:network-discovery}
Before a Digidow transaction as described in section~\ref{sec:architecture:data_flow} can be initiated, the involved parties need to learn about each other. 
Sensors need to know how to contact potential PIAs and the PIAs need to learn how to contact the right verifier.  

In order to achieve this, PIAs take on the additional task of using their information about their owners to predict which sensors an individual might approach in the near future and contacting those sensors preemptively. 
This is enabled by a public directory of all available Digidow sensors; we simply call this the \emph{sensor directory}, independently of its specific implementation (cf.\ section~\ref{sec:protocol:sensor-directory}). 
The directory enables PIAs to discover and register with sets of sensors that an individual might approach based on geographic location.
Sensors keep a temporary list of currently registered PIAs that detected physical world interactions---e.g.\ in the form of biometric data readings---should be compared to\footnote{Again without loss of generality, there are different options on how to communicate and match measured sensor readings with biometric templates stored by the respective PIAs, and which parties contribute to computing the match.}.
Additionally, sensors may provide hints about the verifier whose transactions they are supposed to enable. 
Forwarding this information to PIAs where a biometric match was detected gives them a way to contact the right verifiers, if they opt to continue a transaction~\cite{bib:2022:hoeller:phdthesis}. 
Note that the regulation for European Digital Identity (EUDI) systems requires verifiers (called relying parties in the regulation) to publicly register with their contact details as well as attributes they request~\cite{bib:2024:EU-eIDAS2}.
Such a \emph{verifier directory} can directly be used within Digidow, both for providing guidance to PIA policies or user choice in terms of selective disclosure of attributes (which is the main intention of relying party registration in the regulation) and for network discovery of verifiers as an alternative or augmentation to providing verifier details in the sensor directory.
Within the scope of this paper, we focus on the use of a sensor directory for network discovery purposes.

This introduction of a sensor directory, however, creates a new potential linkability issue because requests to the sensor directory reveal which sensors a PIA wants to register to, which in turn provides an estimate on the current physical location of an individual. 
Therefore, queries against the sensor directory must be made at unpredictable times (to prevent them being linked based on time) and always request all sensors in a large area.  
This (in combination with the network privacy measures from section~\ref{sec:challenges:unlinkability:network-privacy}) provides a sufficient level of k-anonymity while keeping implementation complexity low.
Future iterations of the sensor directory are expected to utilize private information retrieval (PIR)~\cite{bib:1998:chor:private-information-retrieval} to provide even stronger anonymity guarantees. 

Aside from its unlinkability benefits, the sensor directory is also essential to achieve globally scalable service discovery for Digidow components. 
In terms of its main functions, it may be considered like a specific form of DNS, but optimized for conveying relevant sensor attributes and in particular physical location based filtering.

\subsection{Location Prediction}
The approach outlined in section~\ref{sec:challenges:network-discovery} requires PIAs to predict which sensors an individual is likely to interact with in the near future. 
Whereas there are some obvious examples like employees leaving their home in the morning being likely on their way to work, Digidow will require a generic approach that works for different individuals with widely different habits in different environments. 
Support for individuals that rarely interact with sensors or a short transaction history make automated predictions difficult given the current state of the art in machine learning. 
Whereas keeping the responsibility of location prediction with each PIA allows them to adapt specifically to the single individual even to the point of using custom machine learning models, the comparatively small amount of location/behavior data per individual still presents a challenge for completely unsupervised learning methods. %
Therefore, we currently assume the PIA to additionally allow individuals to actively inform it about their current position and intentions, e.g.\ through a smartphone companion app, causing it to register to sensors in the vicinity.

\section{Abstract Protocol Definition}
\label{sec:abstract-protocol-definition}
 
The distributed nature of the Digidow architecture necessitates communication between the various components. 
We now explain how these different interactions are expected to work. 

\subsection{Issuing Attributes}
\label{sec:protocol:issuing-attributes}
Before digital identities can be used, they need to be issued and deployed to the correct PIA. 
Within Digidow, two different forms of issuing attributes are distinguished: 
\begin{enumerate*}
\item The initial deployment of \emph{biometric identity attributes} which requires the issuing authority to establish the identity and the biometric information about the identity holder in advance. 
\item \emph{Additional attributes} can then be deployed by verifying the identity of an individual based on the already issued biometric attributes, significantly reducing the complexity of adding new or updating existing non-biometric attributes. 
\end{enumerate*}
Both roles (biometric identity and additional attribute issuers) can be taken on by any organization, limiting the impact of selective denial-of-service attacks (\ref{threat:denial-of-service}) by individual issuing authorities.
Malicious attributes signed by rogue issuing authorities are mitigated by mandatory trust relationships between verifiers and issuing authorities. 

Issued digital attributes are encoded in the well-established format proposed by the W3C for verifiable credentials~\cite{bib:2022:w3c:vc-data-model}. 
Their nomenclature is used to describe how information about digital attributes is exchanged between Digidow devices. 

\subsubsection{Initial Biometric Attribute Deployment}
\label{sec:protocol:issuing-attributes:initial-attributes}
\begin{figure}
    \centering
	\includegraphics[width=0.85\textwidth]{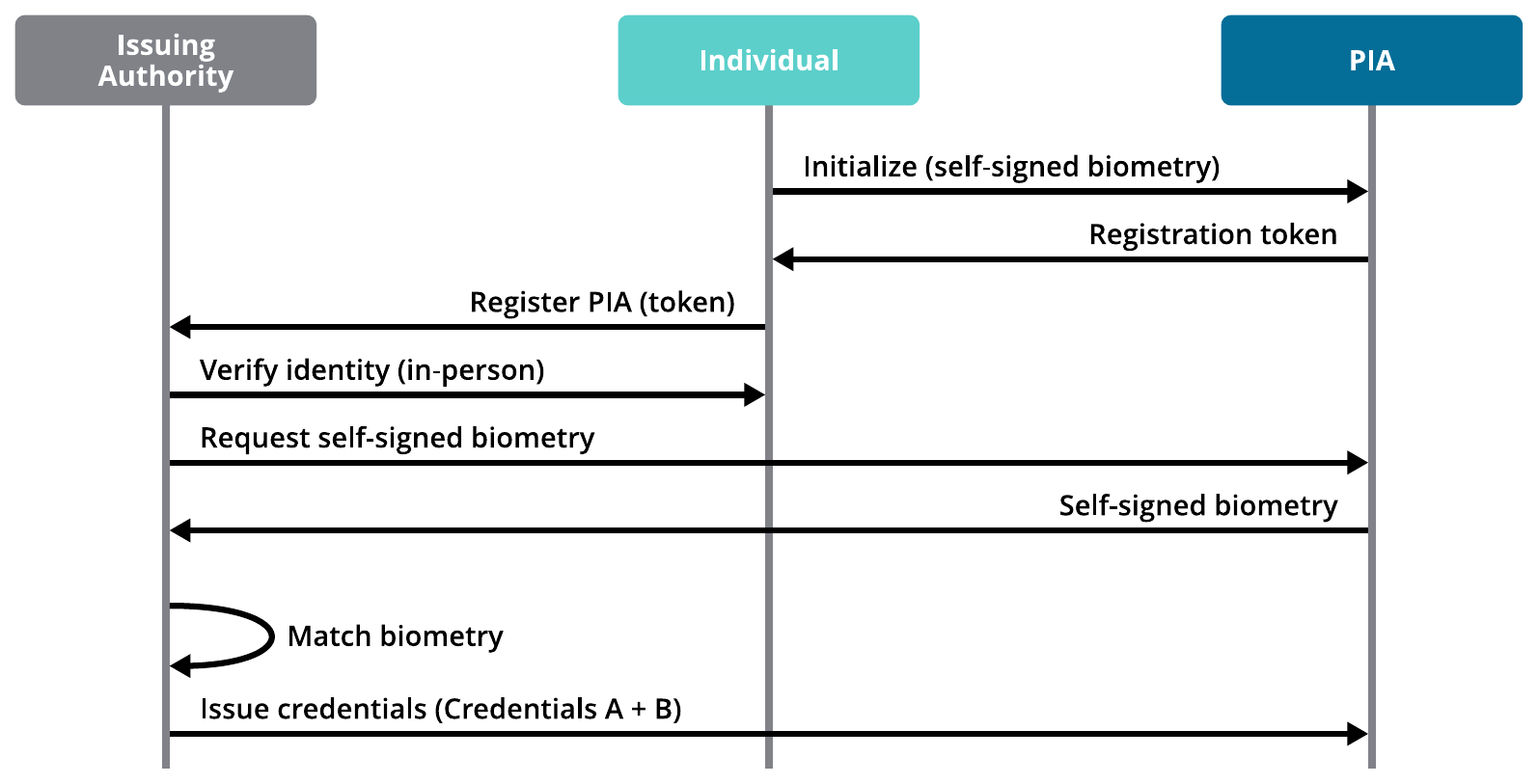}
	\caption{Bootstrapping a PIA}
	\label{fig:protocol:initial_ia}
\end{figure}
Figure~\ref{fig:protocol:initial_ia} shows the process of bootstrapping a PIA by providing it with an initial set of attributes. 
This interaction is responsible for linking an individual to its PIA and usually requires physical presence at the issuing authority. 
It begins with the PIA being initialized by its owner, who provides the PIA with their biometric features and receives a registration token in return. 
This registration token is then shared with the issuing authority, enabling it to contact the PIA and to verify that the individual in front of them has control over the PIA. 
Before a connection to the PIA is established, the issuing authority verifies the individual's identity via alternative means---like checking a traditional physical ID---to prevent identity theft~(\ref{threat:identity-theft}). 
Once the individual's identity has been established, the issuing authority requests the biometric features stored in the PIA and compares them to the biometric features of the requesting individual. 
When they match, the issuing authority can start provisioning verifiable credentials to the PIA. 

\begin{figure}
    \centering
    \subfigure[\label{fig:protocol:credential:a}]{%
    	\includegraphics[width=.40\textwidth]{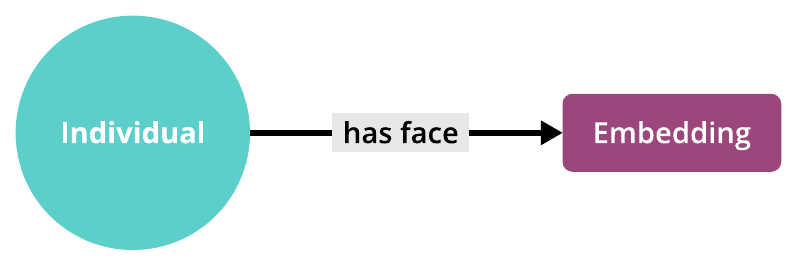}%
    }%
    \qquad%
    \subfigure[\label{fig:protocol:credential:b}]{%
        \includegraphics[width=.40\textwidth]{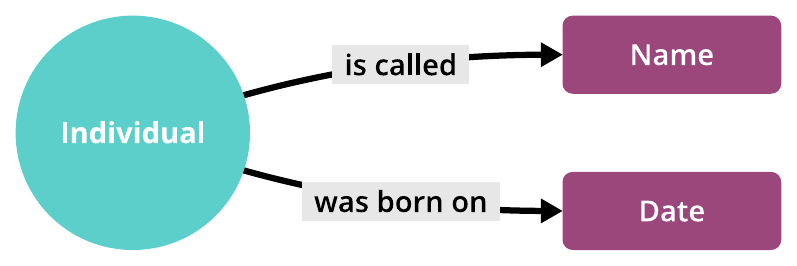}%
    }%
	\caption{Claims contained within \subref{fig:protocol:credential:a}~credential A and \subref{fig:protocol:credential:b}~credential B}
	\label{fig:protocol:credentials}
\end{figure}
Verifiable credentials always contain one or more \emph{claims}; with every claim consisting of a subject, a property, and a value. 
Figure~\ref{fig:protocol:credential:a} shows how such a claim could be used to link biometric measurements (usually in the form of an embedding vector from a deep neural network) to an individual (we call this the \emph{credential~A}). 
The most interesting issue to discuss at this point is what to use as the subject for claims about the individual. 
One option would be to use the public key of a public/private key-pair where only the PIA knows the private key and which has explicit non-transferability properties (e.g., through all-or-nothing sharing~\cite{bib:2001:camenisch:non-transferable-ac}).
Whereas this would work fine for a PIA to prove ownership of an identity, it requires issuing authorities to share all public keys for which they have issued a biometric credential to prevent identity modification (\ref{threat:identity-modification}) by two individuals sharing their PIA's private keys to trick an issuing authority into signing attributes of different identity owners using the same public key. 
Digidow's solution for this issue relies on key derivation by requiring every authority issuing biometric credentials to use a key derivation scheme that turns the client-provided public key into an issuing authority specific pseudonym. 
Every pseudonym can be directly attributed to one specific issuing authority, and it is the responsibility of said issuing authority to ensure that it is never tied to the biometric features of more than one individual. 
The same pseudonym can now be used as subject for all other attributes the issuing authority knows about this individual (see Fig.~\ref{fig:protocol:credential:b}; we call these non-biometric attribute sets \emph{credential~B}). 
Once the PIA has been bootstrapped, updating or obtaining new credentials from the same issuing authority no longer requires physical presence of the individual. 
The PIA can just remotely prove ownership of an identity key to identify itself and then obtain further attributes. 

\subsubsection{Additional Attributes}
\begin{figure}
    \centering
	\includegraphics[width=0.85\textwidth]{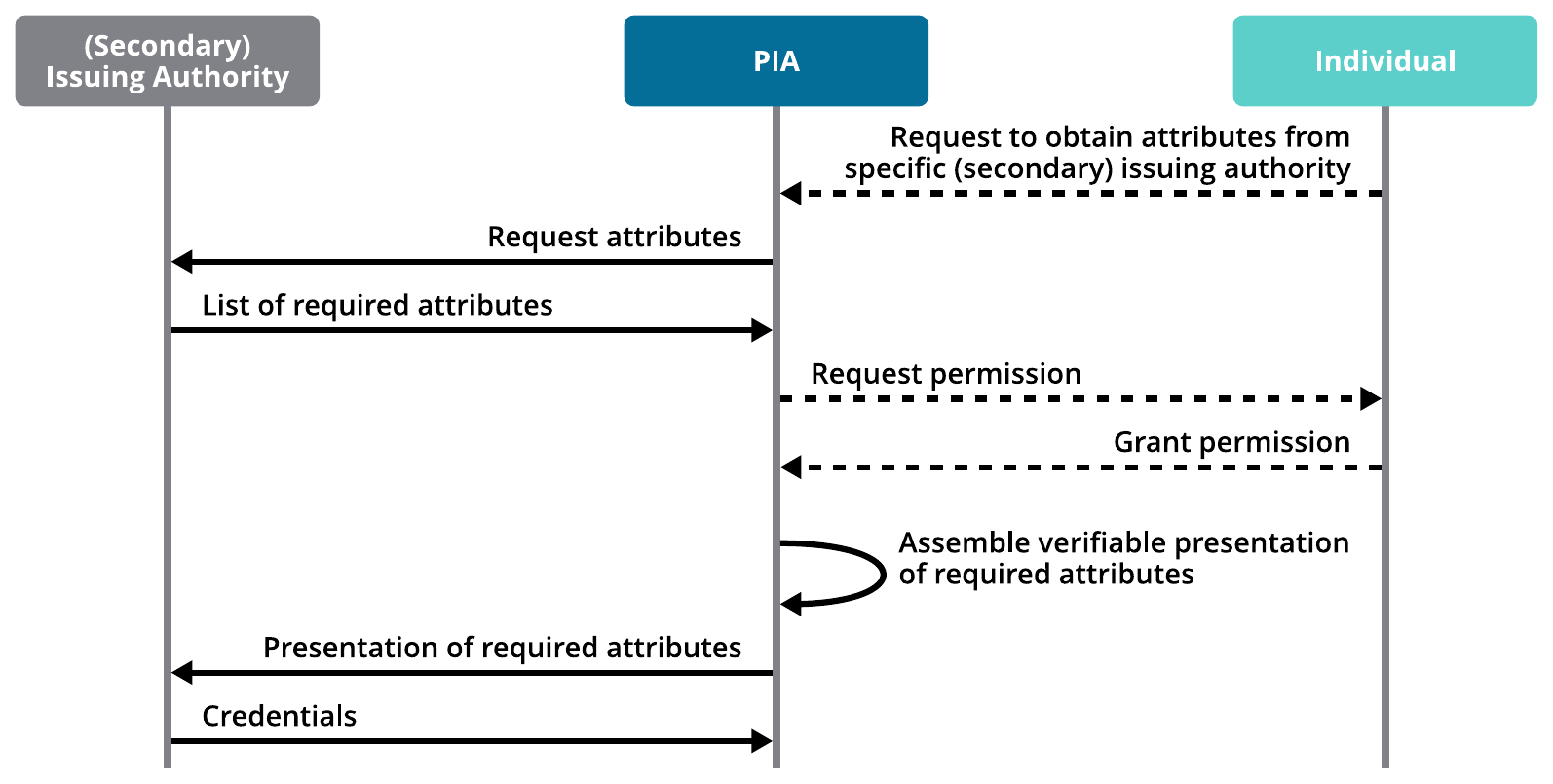}
	\caption{Obtaining (or renewing) additional attributes}
	\label{fig:protocol:additional_ia}
\end{figure}
Once a PIA has been bootstrapped with an initial set of attributes, obtaining or renewing additional attributes from other issuing authorities follows a simplified procedure (see Fig.~\ref{fig:protocol:additional_ia}). 
The interaction can either be triggered by an individual manually or automatically if currently stored credentials are about to expire. 
In both cases, the PIA connects to an issuing authority and asks it to provide it with new credentials. 
The (secondary/additional) issuing authority should respond with a list of attributes it needs to identify the individual along with a set of trusted (initial/biometry) issuing authorities. 
This enables the PIA to find out if it can obtain attributes from this issuing authority. 
If the PIA has the necessary attributes, it now needs to decide if it should disclose these attributes to the issuing authority. 
Depending on the configuration of the PIA, this decision might be taken automatically, or it might request confirmation from the individual before proceeding. 
Once the PIA is certain that sharing the requested attributes with the issuing authority is intended by its owner, it creates a new verifiable presentation---a structure that allows multiple verifiable credentials from different issuers---with all the requested attributes and sends it to the issuing authority. 
This enables the issuing authority to tie the identity key used as subject in the claims within the verifiable credentials to user information within their internal systems. 
Once that link has been established, the issuing authority can now generate new verifiable credentials (of type \emph{credential~B}) containing the information the issuing authority has about the individual. 
Sending these new credentials back to the PIA concludes this interaction. 

\subsection{Using the Sensor Directory}
\label{sec:protocol:sensor-directory}
\begin{figure}
    \centering
	\subfigure[\label{fig:protocol:sensor-directory-sequence:upload}]{%
        \includegraphics[width=.44\textwidth]{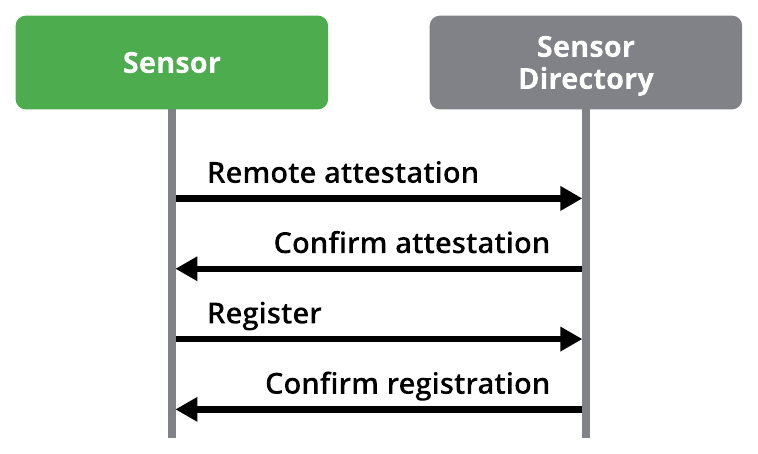}%
    }%
    \hfill%
	\subfigure[\label{fig:protocol:sensor-directory-sequence:download}]{%
        \includegraphics[width=.44\textwidth]{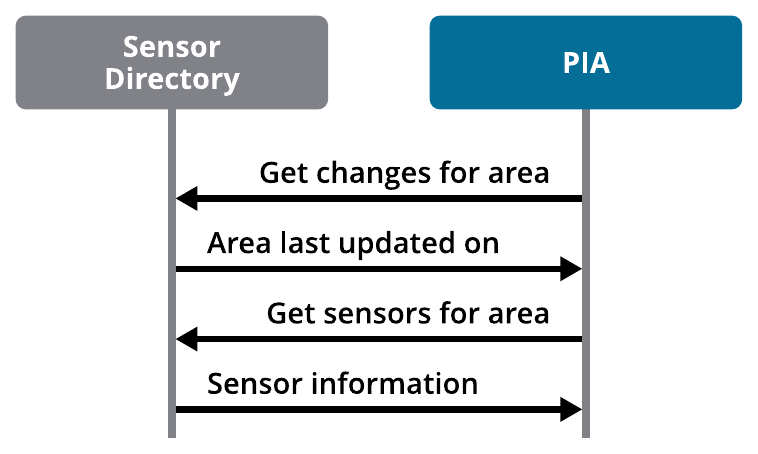}%
    }%
	\caption{Uploading to~\subref{fig:protocol:sensor-directory-sequence:upload} and downloading from~\subref{fig:protocol:sensor-directory-sequence:download} the sensor directory}
	\label{fig:protocol:sensor-directory-sequence}
\end{figure}
The sensor directory is involved in two different interactions: self-registration by sensors (cf.\ Fig.~\ref{fig:protocol:sensor-directory-sequence:upload}) and retrieval of published sensor information by PIAs (cf.\ Fig.~\ref{fig:protocol:sensor-directory-sequence:download}). 

Publishing a sensor is a fairly simple interaction: 
First, the sensor remotely proves that it is a properly configured Digidow sensor using its internal RoT. 
This ensures that only valid sensors end up in the sensor directory and makes it harder for malicious actors to flood the sensor directory with bogus entries. 
Once this remote attestation is verified, the sensor may publish data about itself, such as sensor type and location. 
Every sensor---or more precisely every RoT---is only allowed to publish one entry, so a second registration updates the existing sensor instead of adding a new one. 

Obtaining information about available sensors requires more attention because neither the sensor directory nor passive network observers should be able to identify a PIA based on the sensors it is interested in (\ref{threat:data-collection}). 
Whereas it is expected that a PIR based approach will solve this issue in the future, it is temporarily mitigated by having a PIA always request all sensors within a defined area. 
Digidow proposes the use of ZIP codes as areas because they divide regions based on geography and population, supposedly resulting in comparatively similar numbers of downloads to make correlation attacks less probable. 
Additionally, a PIA locally caches the information obtained from the sensor directory (it can even add private sensors to the internal list) and refreshes them at regular yet unpredictable (i.e., randomized) intervals. 
Therefore, the interaction for obtaining new sensors begins with the PIA checking if the sensor directory has new sensors within a certain area. 
Only if there is something new to be obtained, the PIA downloads the new sensor data batch.

\subsection{Digidow Transactions}
\label{sec:protocol:transaction}
\begin{figure}
    \centering
    \subfigure[\label{fig:protocol:transaction-sequence:identification}]{%
        \includegraphics[width=0.44\textwidth]{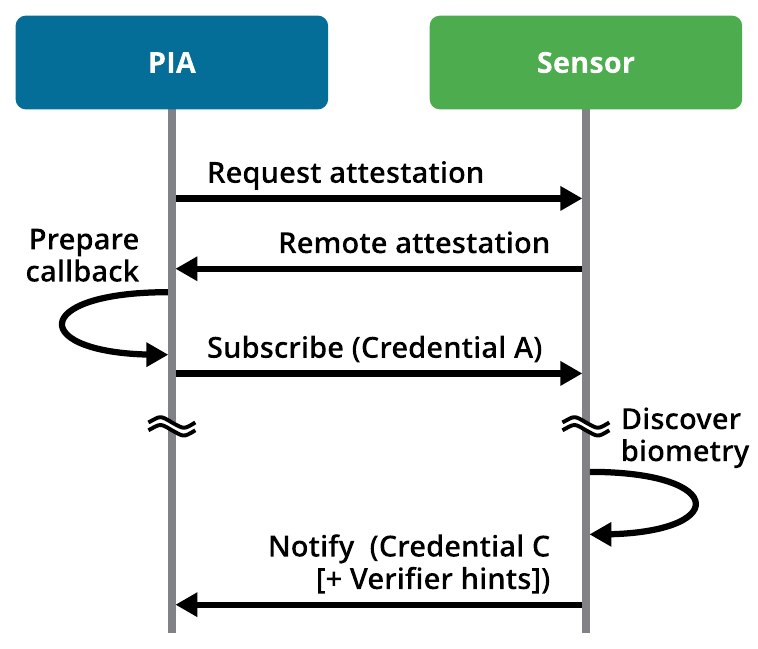}%
    }%
    \qquad%
    \subfigure[\label{fig:protocol:transaction-sequence:identification-credential}]{%
        \includegraphics[width=0.44\textwidth]{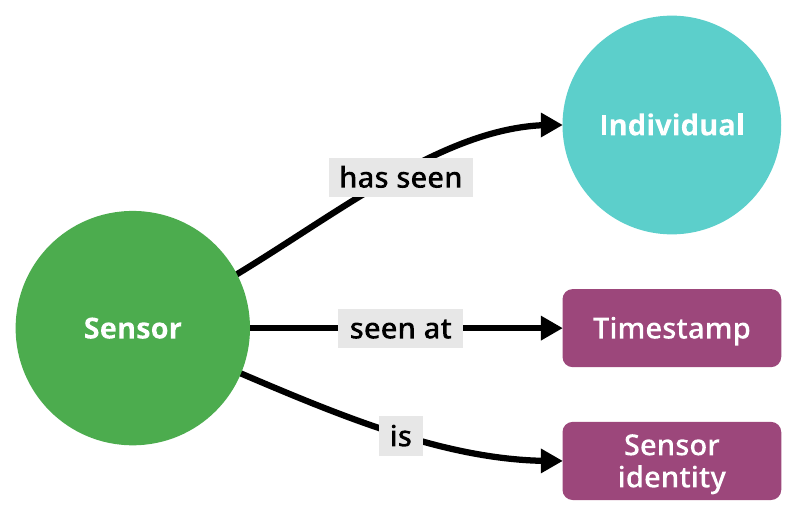}%
    }%

    \subfigure[\label{fig:protocol:transaction-sequence:authentication}]{%
        \includegraphics[width=0.85\textwidth]{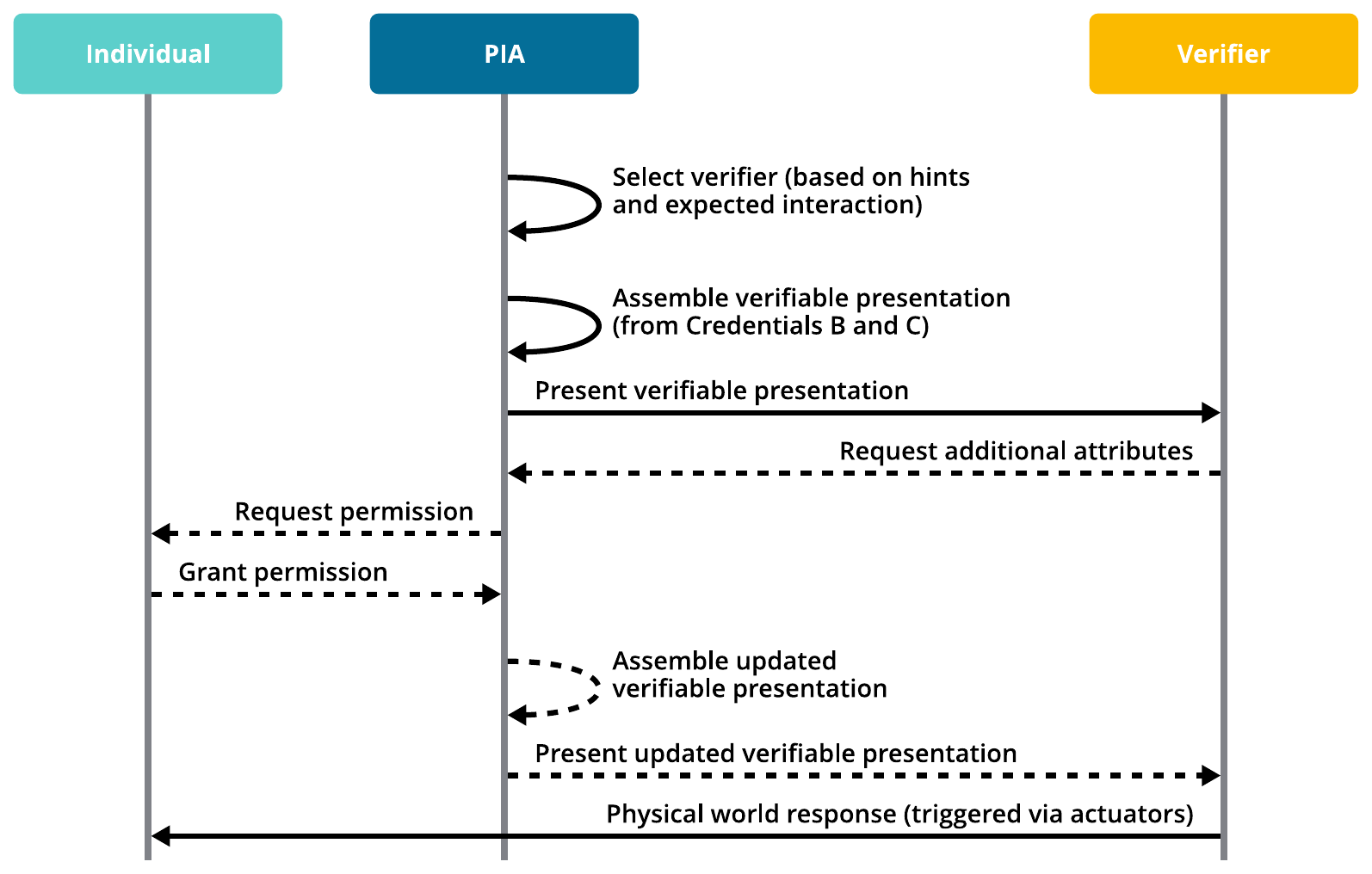}%
    }%
    \caption{Interactions during a Digidow transaction: \subref{fig:protocol:transaction-sequence:identification} identification by a sensor, \subref{fig:protocol:transaction-sequence:identification-credential} credential C with attesting sensor identification, and \subref{fig:protocol:transaction-sequence:authentication} authorization at a corresponding verifier}
    \label{fig:protocol:transaction-sequence}
\end{figure}
A standard Digidow transaction, in which digital identities are used to trigger actions in the physical world, consists of two different interactions (cf.\ Fig.~\ref{fig:protocol:transaction-sequence}): 
The sensor and the PIA communicate first to establish that a specific individual has been detected by a sensor at a specific physical location. 
Afterwards, the PIA can engage a verifier to trigger the appropriate response to its owner's presence. 

The first interaction begins when a PIA predicts that its owner might interact with a specific sensor in the near future. 
This causes the PIA to establish trust in the sensor by verifying its current hardware and software configuration. 
If the PIA considers the sensor trustworthy, it registers with an ephemeral callback address (\ref{threat:data-collection}, \ref{threat:denial-of-service}), a registration duration, biometric comparison data, and a \emph{credential~A} proving that this PIA is owned by the individual with the provided biometric data (mitigating \ref{threat:data-collection} by preventing registration of arbitrary biometrics). 
This biometric comparison data depends on the type of biometry the sensor understands and the type of matching algorithm the sensor employs. 
If possible, the comparison should be privacy preserving and not require the PIA to transmit their full biometric template.
However, if the PIA has established sufficient trust that the sensor will neither store nor leak the template, even sharing the template directly with the sensor is an option. 
Since a PIA will not be able to accurately predict the next actions of its owner in the general case, PIAs are expected to subscribe to many sensors in parallel to ensure that they are always registered to sensors before their owner interacts with them. 

Once the registration has been completed, the remaining part of the transaction is triggered when the individual interacts with one of these pre-registered sensors. 
At this point, timing requirements of the interaction also change significantly---now an individual is waiting for a response, whereas everything until this point has been happening ahead of time. 
After a sensor detected a match, it uses the callback address specified in the subscription request to provide the PIA with a new \emph{credential C} confirming that the identity key---for which the matching biometric was issued (mitigates \ref{threat:identity-theft})---was detected by the sensor at a specific time (see Fig.~\ref{fig:protocol:transaction-sequence:identification-credential}). 
In addition to the credential, a sensor can be configured to provide the PIA with hints about verifiers that might be interested in the credential. 
In most cases, sensors will be deployed for specific verifiers and providing PIAs with information about them ensures fast transactions for users. 
The callback information is then used by the PIA to update its embedded location model responsible for predicting its owner's physical location and to estimate if its owner intends it to contact a verifier. 
We expect that the history of previous transactions along with the verifier hints will typically allow the PIA to make this decision itself.
However, for some cases (for example using a sensor which does not provide verifier hints for the first time) the PIA will be required to reach out to additional information sources such as potential verifier directories (cf.\ section~\ref{sec:challenges:network-discovery}) or contact its owner for further explicit instructions. 

Then, the PIA uses the information from previous transactions or the verifier hints to determine which attributes are required by the verifier, and constructs a \emph{verifiable presentation} containing the verifiable \emph{credential~C} received from the sensor (mitigates \ref{threat:identity-modification}) along with additional \emph{credentials~B} from other issuing authorities for only the attributes the verifier is interested in (mitigates \ref{threat:data-oversharing}). 
Sending this presentation to the verifier should cause the verifier to communicate with its actuators to trigger the physical world response that the PIA owner is waiting for. 

If a verifier finds the attributes presented by the PIA to be insufficient, it can respond to the PIA asking for additional attributes. 
This will most likely result in the PIA contacting its owner to inform them that a Digidow transaction has failed, and that additional attributes have been requested. 
If the PIA has the requested attributes, the owner is given the choice of authorizing the transfer of additional attributes to the verifier. 
If they are not yet issued to the PIA, the owner has to instruct the PIA to obtain them from an appropriate issuing authority before the transaction can be completed successfully. 
Once the PIA has the necessary attributes and the permission to forward them, a new verifiable presentation can be constructed and forwarded, concluding the interaction between the PIA and the verifier.

\section{Evaluation}
\label{sec:prototype}
We now demonstrate that a digital identity system based on the Digidow architecture (section~\ref{sec:architecture}) following the abstract protocol proposed in section~\ref{sec:abstract-protocol-definition} can be implemented with currently available technologies.
Additionally, we present an initial baseline assessment of its efficiency.
We focus on the most challenging aspects outlined in section~\ref{sec:challenges} in our implementation of a Digidow prototype and, in the following sections, explain how current state of the art provides the means to address them. 
Beyond these sufficiently well-understood methods and software engineering aspects of their practical application, we highlight a few particular areas of ongoing research.
Whereas we already see significant results in these areas, we still consider them potentially open for fundamental improvement based on future results in the respective sub-fields.

\subsection{Network Privacy}
\label{sec:prototype:network-privacy}
Preventing the various interactions presented in section~\ref{sec:abstract-protocol-definition} from leaking personal information about users on the network communication layer primarily requires concealing metadata about who is communicating with whom (cf.\ section~\ref{sec:challenges:unlinkability:network-privacy}). 
Whereas multiple systems---for example Tor~\cite{bib:2004:dingledine:tor,bib:tor-project}, I2P~\cite{bib:i2p}, or Nym~\cite{bib:2021:diaz:nym-whitepaper}---provide a reliable way to obscure metadata, only the Tor network has at the time of this writing seen enough usage, attacks, and continuous improvements to reach the level of maturity required by digital identity systems~\cite{bib:2022:hoeller:phdthesis}. 

The Tor network provides clients with a way to anonymously accept incoming connections (onion services~\cite{bib:2023:tor:rendezvous-spec-v3}) as well as to anonymously make outgoing connections, but there are two significant disadvantages: 
First, the individuals behind an onion service remain anonymous, but it is still possible to track if and when an onion service is being contacted. This implies that a PIA cannot use the same onion address as callback address for all subscriptions without leaking linkable information about when its owner interacts with a sensor~\cite{bib:2021:hoeller:foci}. 
Second, onion services require multiple connections across several different circuits before they are ready for communication. 
Building circuits to the hidden service directory, an introduction point, and a rendezvous point, along with the time needed to set up a working connection via these circuits, takes significantly longer and consumes more resources than a regular direct connection would. 

The first issue can be addressed by using a new onion service and onion address for every subscription. 
If every onion address is only used once and there is no way to determine if two onion services belong to the same PIA, there is no point in observing if and when it is being created and contacted (mitigating  \ref{threat:data-collection} and  \ref{threat:denial-of-service}).
However, onion services are not intended to be set up on the fly and trying to do so adds several additional seconds to every Digidow subscription request~\cite{bib:2021:hoeller:wtmc}. 
Fortunately, this issue can be avoided if PIAs maintain a set of working onion services that have not been used yet. 
However, this forces a PIA to maintain a significant amount of onion services, both those currently in use and those intended for future use. 
Every running onion service requires at least one dedicated active connection through the Tor network, so operating multiple onion services in parallel causes a linear increase in spent networking and computing resources. 
This increases the hardware capabilities required by PIAs and makes it infeasible to run a PIA on battery powered devices like smartphones~\cite{bib:2017:kollmann:cost-of-push-notifications-via-tor}. 
Even PIAs with sufficiently powerful hardware and no power restrictions are limited by the Tor network itself, because the Tor client never uses a single relay for more than one simultaneous connection. 
Once a Tor client is using every relay for one connection, it is unable to create further connections, independent of available hardware resources.

Despite these obstacles, achieving a sufficient level of network privacy for a digital identity system (mitigating \ref{threat:data-collection}) is possible. 
The challenge is doing so while also retaining acceptable performance. 
Whereas Tor normally introduces performance penalties, significant improvements are possible due to the fact that regular onion services are designed to handle arbitrary TCP connections, whereas the Digidow architecture has a very limited set of messages and message types.
This enables certain shortcuts compared to regular onion services. 
First, it is possible to use onion services without involving the hidden service directory by directly exchanging service descriptors between PIA and sensors. 
Service descriptors are usually exchanged indirectly through the public hidden service directory because they need to be renewed regularly. 
For a PIA subscribing to a sensor, the limited lifetime is not an issue because the subscription timeout can be expected to be much shorter than the 24-hour lifetime of a service descriptor. 
Not having to publish service descriptors to the hidden service directory reduces the time needed to create onion services by approx.~80\,\% and eliminates the 16 Tor circuits usually needed to publish a service descriptor~\cite{bib:2021:hoeller:wtmc}. 

Another potential improvement can be found when realizing that many Digidow interactions do not require a bidirectional communication channel. 
Subscribing to a sensor, notifying a PIA, or providing a presentation to a verifier do not require an immediate response and, therefore, have no need for a response channel. 
Regular onion service connections start with an introduction cell instructing the onion service to establish a bidirectional communication channel via a rendezvous point. 
If there is no need for such a channel, a client can replace the introduction cell with a data cell directly containing the one-way message itself~\cite{bib:2022:hoeller:phdthesis}.
To prevent passive adversaries from noticing this kind of communication, the data cell must have the same size as the usual introduction cell, limiting the message size to 512~bytes%
, which is insufficient for service descriptors.
If a service descriptor is directly exchanged with the intended recipient without involving third parties, a lot of padding, encryption, and signatures can be stripped without security impact. 
Ultimately, only 96~bytes of data, which easily fit into a cell, are actually needed to connect to an onion service~\cite{bib:2022:hoeller:phdthesis}.

Combining all these approaches reduces the average time needed for a Digidow transaction (subscription + notification) from more than 19~seconds down to only 5.4~seconds~\cite{bib:2022:hoeller:phdthesis}.
This promising intermediary result supports the argument that the network anonymity and performance required by a distributed digital identity system can be achieved by continuous improvements of existing technologies.  

\subsection{Hardware Root-of-Trust}
\label{sec:prototype:hw-rot}
Some security and privacy guarantees required in the Digidow architecture can be achieved through cryptographic or organizational (i.e., separation of concern and compartmentalization between different administrative realms of control) means.
However, some components---in particular the sensors embedded ubiquitously into the environment and processing highly sensitive biometric data---need to be granted a significant level of trust (cf.\ section~\ref{sec:challenges:unlinkability:trusting-sensors}).
PIAs and verifiers also handle sensitive data, but their respective participation in the protocol flow is aligned with their own interests; sensors, on the other hand, need to act as proxies to keep the security and privacy guarantees of stakeholders other than those that directly operate them.
Whereas the direct attack surface can be limited in the sense of sensor or network input targeting these embedded devices themselves, individuals interacting with sensors (or simply standing within their sensing range, without even actively participating in the system) have no choice but to trust that their behavior is according to the specification.
This creates a difficult situation: sensors are operated by a stakeholder (e.g., the organization also running verifiers or an independent entity) whose inherent interests are not aligned with the users who are in direct interaction with them.
The resulting power imbalance needs to be mitigated so that individuals can realistically trust that they can still retain control over their personal data, including biometric attributes (mitigating \ref{threat:identity-theft}, \ref{threat:data-collection}, \ref{threat:denial-of-service}).
 
Whereas cryptographic techniques such as multi-party computation (MPC) for determining a match between biometric templates kept at PIAs and current sensor readings processed at sensors (section~\ref{sec:challenges:biometric-privacy}), fully homomorphic encryption, or zero knowledge proofs can all mitigate the data leakage caused through protocol interactions with other Digidow components, limiting their own misbehavior against, e.g., simply storing all detected biometric data, requires embedded parts that are not under the full control of even the owner respectively operator of the sensor. Such embedded parts that intentionally limit what a device can do are typically hardware based, and their design and implementation are usually certified by independent third parties.

Current state of the art includes different variants of such hardware roots of trust: 
TPMs~\cite{bib:2019:tcg:tpm} and embedded secure elements (SE, e.g., (e)SIM, NFC companion smart card chips or dedicated key store hardware in current smartphones\footnote{For Android devices, dedicated smart card hardware for key storage is called StrongBox.}) are the simplest and therefore supposedly most secure form reaching the highest levels of security certification, but limited in features and performance; 
Trusted Execution Environments (TEE) such as ARM TrustZone, Intel SGX, or AMD SEV offer the full level of performance and flexibility of the main CPU, but are complex and have been subject to many different attacks in the past. 
For embedded sensors with limited system complexity but high levels of required trust, we strongly prefer simple RoTs like TPMs or SEs. 

In addition to these dedicated hardware based RoTs, the main CPU needs to have a secure primary bootloader that limits---and ideally measures and communicates to the RoT---which operating system bootloaders can be loaded. %
Each following stage in the boot process is then responsible for ensuring the same guarantees for the respective next stage, i.e., the OS bootloader verifies and measures the OS kernel, the kernel its hardware drivers and system image, etc.

In combination, existing state of the art in secure/verified/measured boot coupled with hardware RoT elements and remote attestation protocols like DAA~\cite{bib:2004:brickell:daa}, DICE~\cite{bib:2018:tcg:dice,bib:2021:tao:dice} or simpler batched-key schemes used with earlier Android key attestation~\cite{bib:aosp-key-attestation,bib:android-cdd-14} is already capable of verifying which system image has been booted and allowing remote verification of this boot state. 
A practical, well-documented, and demonstrably massively scalable example is Android Verified Boot (AVB)~\cite{bib:aosp-verified-boot-2.0-spec} together with key attestation~\cite{bib:aosp-key-attestation} from StrongBox SE-based or KeyMint TEE-based RoTs. 
With a locked bootloader, Android apps can use hardware bound private keys for digital signatures with a key attestation certificate provided by the RoT that includes the bootloader state and, in modern versions, the top-level hash of the booted system image as measured by the bootloader at boot time~\cite{bib:2021:mayrhofer:android-platform-security-model}. 
We argue that these exact same methods can be used---potentially by directly using a read-only Android Open Source Project (AOSP) base system built reproducibly~\cite{bib:2022:poell:wisec} and executed on standard ARM CPUs with existing bootloaders and open source TEEs like Trusty or combined with off-the-shelf TPM modules---to achieve strong trust in the boot-time state of Digidow sensors.

However, there are currently two open shortcomings with this approach: 
\begin{enumerate}
	\item Verification of RoT certification, i.e., the roots of trust for verifying the embedded hardware RoT elements themselves is inherently difficult in a decentralized system that does not assume a single or only a few main device vendors (we refer to, e.g., \cite{bib:2018:maene:hardware-based-trusted-computing,bib:2021:johnson:embedded-remote-attestation-schemes} for an overview of the diversity of hardware RoTs). 
	\todoI{skipping for #SPACE reasons}
	On top of the difficulty of verifying RoT certificates, we need to expect a significantly larger number of different system images attested by these hardware RoTs%
	\todoI{skipping for #SPACE reasons}
	due to various legitimate changes over time that need to be recognized by PIAs and verifiers.
	To avoid depending on large central databases, we propose to add additional metadata to the sensor directory for decentralized RoT certificate and image hash databases%
	---e.g., referring to transparency logs as a mitigation against insider attacks.
	\todoI{skipping for #SPACE reasons}

	\item Attestation stops at boot time; run-time integrity of active systems depends on the assumption of secure, near-perfect code---free of run-time vulnerabilities that could be exploited after boot-time system image measurement.
	Previous experience\footnote{Statistics of CVEs reveal a continuous increase in the number of discovered software vulnerabilities over the past years, cf.\ \url{https://www.cvedetails.com/vulnerabilities-by-types.php}.} strongly indicates that this assumption is (at least with the current state of secure software engineering) wrong, and boot-time measurements provide limited guarantees of run-time security.
	\todoI{skipping for #SPACE reasons}
	None of the currently deployed approaches offer any form of run-time inspection and attestation of the dynamic in-memory state of these user space processes.%
\end{enumerate}
Our research prototype therefore focuses on the orthogonal issue of reducing both variability and run-time complexity of sensor system image code while embedding state-of-the-art machine learning models, e.g., for live face detection and recognition with deep neural network embeddings~\cite{bib:2023:hofer:momm,bib:2023:hofer:aaiml}.
Further minimizing both boot-time and run-time code dependencies, e.g., by compiling the main sensor pipeline together with an unikernel for decreased attack surface and variability, is subject to future work, but already supported by using Rust and its build tooling for bare-metal binary execution.
Combining such minimized sensor pipeline code with the existing state-of-the-art for secure boot-time verification and remote attestation is, arguably, currently the best approach for future practical deployment of trustworthy sensors.
 
\subsection{Cryptography}
\label{sec:prototype:cryptography}
One central goal of any identity system is to prevent identity theft (\ref{threat:identity-theft}), which requires preventing forgery of attributes or transfer of attributes to different individuals (\ref{threat:identity-modification}).
To provide these \emph{anti-forgery} guarantees, issuing authorities need to include a proof of integrity and authenticity with all attributes they provide to individuals respectively their PIAs. 
The simplest cryptographic form of such proof is a digital signature with a private key held by the issuing authority.
If attributes are signed individually rather than the complete set of attributes contained in a credential document, they can also be presented individually, allowing for selective disclosure (as described in section~\ref{sec:challenges:cryptographic-privacy}).
However, binding attributes to the underlying individual identity---and their biometric templates---is required for effective anti-forgery~\cite{bib:2024:garciarodriguez:privacy-preserving-access-control}.
Signatures therefore need to associate attributes with the (biometric) identity while providing selective attribute disclosure (mitigating \ref{threat:data-oversharing}) and unlinkability (mitigating \ref{threat:data-collection}) not only in attributes transmitted to a verifier, but also for these items of proof \emph{even if issuing authorities and verifiers collude} (as outlined in section~\ref{sec:challenges:unlinkability}).

Traditional digital signature schemes do not provide unlinkability, as each signature block attached to an attribute (or more specifically, the binding of an attribute to an identity) is easily linkable between different presentations. 
A workaround is that issuing authorities provision many different signature blocks for (sets of) attributes, and presenting each unique signature only once. 
The standard for mobile driving licenses, ISO/IEC 18013-5, and the generic ISO/IEC 23220 mobile eID standards series recommend (but do not require) so-called single-use mobile security objects (MSO, cf.\ section~\ref{sec:comparison:mdl}), which---through an indirection of hash sets---allow selective disclosure together with attribute binding for exactly this purpose of providing unlinkability of signatures~\cite{bib:2021:iso:18013-5,bib:2023:iso:23220-1}.
There are three obvious disadvantages of this approach:
\begin{enumerate}
	\item Credentials increase in size, linearly with the maximum number of (unlinkable) presentations.
	\item PIAs need to regularly receive new attribute signature blocks, which is practically similar to re-provisioning new credentials, and therefore hinders offline scenarios and leaks credential usage information back to the issuing authority.
	\item Critically, whereas such single-use signatures are unlinkable to (collusion sets of) verifiers, collusion with issuing authorities trivially breaks unlinkability, because issuing authorities directly provide each single-use signature and can therefore easily associate them to individual identities (cf.\ \cite{bib:2024:arf-feedback} for an argument why this is in violation of the eIDAS regulation~\cite{bib:2014:EU-eIDAS}).
\end{enumerate}
Non-interactive zero-knowledge proof (NIZKP) schemes such as zk-SNARK~\cite{bib:2019:petkus:zksnark}, zk-STARK~\cite{bib:2019:ben-sasson:zkstark}, or Bulletproofs~\cite{bib:2018:buenz:bulletproofs} seem like ideal building blocks for (offline capable) unlinkable presentation of attributes bound to (biometric) identity.
Unfortunately, their main disadvantages include (in part significantly) larger proof sizes, significantly slower creation and verification times compared to conventional signatures, and---for some schemes---the assumption of a trusted setup process.

Therefore, so-called anonymous credentials (AC)~\cite{bib:1985:chaum:security-without-identification,bib:2000:lysyanskaya:pseudonym-systems,bib:2001:camenisch:non-transferable-ac} (cf.~\cite{bib:2023:kakvi:ac-sok} for a recent survey) describe digital signature schemes that allow selective disclosure of messages and, critically, for which a signature does not break unlinkability; these can be seen as a specific form of zero-knowledge proof for identity systems.
Earlier seminal works have proposed such constructions to support unlinkability of provers (in our case, a PIA presenting certain subsets of attributes to verifiers) without resorting to generic NIZKP schemes. 
The most well-known include BBS~\cite{bib:2004:boneh:bbs,bib:2023:tessaro:revisiting-bbs} and CL~\cite{bib:2004:camenisch:cl} using zero-knowledge proofs of knowledge of signatures over (blinded) multi-element messages (which can, e.g., be tuples of attributes). Following those seminal works, a more recent main class of schemes providing randomizable signatures such as PS~\cite{bib:2016:pointcheval:ps} have been developed, which can loosely be described as a more efficient version of CL with constant-size signatures independently of the number of signed attributes.
Note that the notion of randomizing an existing signature without knowledge of the signing key (which can also be described as self-blinding), which was originally proposed with the CL scheme, is indeed one potential building block for distributed, unlinkable identity systems: 
PIAs can create single-use signatures for every transaction without having to interact with the original issuer, and even collusion between verifiers and issuing authorities does not break unlinkability, because the randomization value is determined by the (non-colluding) PIA. 
Many of these schemes also require the maximum number of elements for message tuples (i.e., the maximum number of attributes in each credential) to be fixed at setup time when creating the issuing authority signing keypairs. 
This is a practical disadvantage in terms of (private and public) key size, but not necessarily for the signature size (e.g., for PS the signatures are constant size).
Additional desirable properties include delegation of credentials and potentially even hiding the issuing authority itself, e.g., to not leak information about the citizenship of an individual simply based on the authority that signed their attributes. 
Although newest results suggest signing methods that support these features~\cite{bib:2023:mir:popets,bib:2023:mir:phdthesis}, they have not yet seen sufficient independent analysis for potential standardization efforts.
For a recent, in-depth review of randomizable key signature schemes that directly support issuer hiding, we refer to~\cite{bib:2023:celi:randomizable-signatures-sok}.

For our prototype implementation, we currently use BBS(+) signatures as implemented in the Rust crate\footnote{\url{https://docs.rs/bbs/}}---at the moment without the desirable delegation or issuer-hiding properties of other schemes~\cite{bib:2023:mir:popets}. 
BBS is still capable of providing selective disclosure and unlinkability properties as long as all involved parties support the required pairing-friendly elliptic curve primitives.
Current IETF RFC efforts include BBS~\cite{bib:2023:draft-irtf-cfrg-bbs-signatures-05}, and it is therefore one potential option for upcoming identity systems, also explicitly recommended in~\cite{bib:2024:arf-feedback}. 

Another recent proposal~\cite{cryptoeprint:2024/2010} applies a NIZKP using highly optimized circuits based on the Ligero and sumcheck systems to the presentation of otherwise unmodified credentials with ECDSA signatures---such as mDL (section~\ref{sec:comparison:mdl}) in already deployed \emph{mdoc} format.
This potentially generic approach to ensuring unlinkability at presentation time does not require a trusted setup, has (compressed) proof sizes in the range of a few 100\,kB, and can generate proofs in less than 2\,s on modern smartphones.
It is currently aiming for standardization under the name \emph{libZK}~\cite{bib:2025:draft-google-cfrg-libzk-00}.
We are further evaluating the current implementation of libZK\footnote{\url{https://google.github.io/longfellow-zk/}}, but Digidow remains open to other signature schemes with the required properties, and can easily support multiple schemes in parallel depending on future standardization directions.

\subsection{Proof of Concept Implementation}
\label{sec:prototype:implementation}
We implement an initial, proof of concept prototype that combines the most important research results into a single, consistent experimental environment.
To ensure that our experiments remain aligned with real world scenarios, they were conducted with methods and technologies that would also be suitable for real-world use within a digital identity system dependent on user trust. 
Rust is used as a programming language that allows for secure, efficient, and reproducible code. 
We unify codebases where reasonably possible, e.g.\ through a Rust library implementing the sensor-side Digidow protocol flow that is being used as the basis of all prototype sensors\footnote{\url{https://git.ins.jku.at/proj/digidow/sensor-lib}}, including an optimized face recognition pipeline\footnote{\url{https://git.ins.jku.at/proj/digidow/face-lib}} executed on embedded edge devices~\cite{bib:2023:hofer:momm,bib:2023:hofer:aaiml}.
Similarly, a single Rust codebase implements the PIA-side components and can be executed both in thin cloud services (currently packaged as a single self-contained binary as well as a Docker container image) and embedded in Android apps~\cite{bib:2022:poell:masterthesis}.
Nix~\cite{bib:2006:dolstra:nix} was chosen as a software-defined reproducible build environment to support transparency of build and deployment processes. 
Experiments are built within a continuous integration pipeline that verifies reproducibility of created artifacts~\cite{bib:2022:schwaighofer:nixcon}. 
Successfully built artifacts are added to transparency logs to mitigate risks of unauthorized usage of the signing key (e.g., through insider attacks)~\cite{bib:2023-lins-nordsec}. 

While this prototype implementation should not yet be considered for practical use, it demonstrates how all involved concepts can be combined and enables initial measurements to validate our design choices on multiple layers: end-to-end system performance testing, targeted component level assessment to identify bottlenecks, and initial scalability estimation.

\subsubsection{End-to-end system performance}

Our first validation approach measures the total latency from facial detection to the verifier's receipt of the verifiable presentation required for action authorization.
This test traces the complete workflow through three system components:
\begin{enumerate}
	\item The sensor\footnote{\url{https://git.ins.jku.at/proj/digidow/sensor}} captures and processes camera images, performs feature mapping from 2D to 3D with distortion correction, detects and aligns faces, recognizes faces based on current registrations and specified thresholds, calculates distance parameters, creates the verifiable presentation, and transmits it to the PIA.
	\item The PIA\footnote{\url{https://git.ins.jku.at/proj/digidow/pia}} records this transaction, verifies the presentation and its plausibility, identifies the sensor, creates a new presentation based on the sensor data, and forwards it to the verifier.
	\item The verifier\footnote{\url{https://git.ins.jku.at/proj/digidow/verifier-door}} then validates the sensor credentials, extracts key data such as door ID and matches it to internal mappings, confirms validity timing, checks cool-down periods, verifies the PIA's credentials, and finally transmits the door unlock command.
\end{enumerate}

\begin{figure}
    \centering
    \subfigure[\ldots\ communication over local connection \label{fig:prototype:endtoendmeasurement:local}]
    {%
        \begin{tikzpicture}[font=\small]
            \begin{axis}[
                ybar,
                xlabel={Seconds},
                ylabel={Frequency [samples]},
                ylabel style={anchor=north},
                title={Histogram with local communication},
                ymin=0,
                bar width=10pt,
                xticklabel style={rotate=45, anchor=east},
                enlarge x limits=0.05,
                width=0.48\textwidth, %
                height=0.34\textwidth,
            ]
            \addplot+[hist={bins=10}] table [y index=0] {data/end-local.csv};
            \end{axis}
        \end{tikzpicture}
    }%
    \hfill
    \subfigure[\ldots\ communication over Tor \label{fig:prototype:endtoendmeasurement:tor}]
    {%
        \begin{tikzpicture}[font=\small]
            \begin{axis}[
                ybar,
                xlabel={Seconds},
                ylabel={Frequency [samples]},
                ylabel style={anchor=north},
                title={Histogram with Tor communication},
                ymin=0,
                bar width=10pt,
                xticklabel style={rotate=45, anchor=east},
                enlarge x limits=0.05,
                width=0.48\textwidth, %
                height=0.34\textwidth,
            ]
            \addplot+[hist={bins=10}] table [y index=0] {data/end-tor.csv};
            \end{axis}
        \end{tikzpicture}
    }%
	\caption{Average over 400 samples of end-to-end runtime performance \ldots}
    \label{fig:prototype:endtoendmeasurement}
\end{figure}
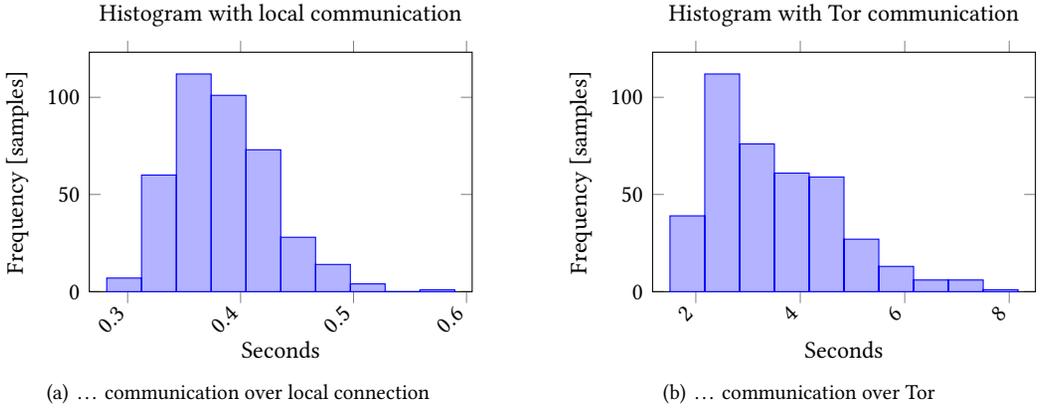

For these measurements, we established all components locally on a laptop with an AMD Ryzen~7 7840U CPU, with the sensor deployed on a Raspberry Pi 5 equipped with an AI hat\footnote{\url{https://www.raspberrypi.com/products/ai-hat/}}.
We conducted tests both with the live Tor network and without Tor by proxying requests over local TCP sockets.
As shown in Fig.~\ref{fig:prototype:endtoendmeasurement}, mean processing time for local connections was 384 ms, while the Tor-based communication averaged 3495 ms\footnote{Note to reviewer: The Tor data sample was collected between 2025-03-21 and 2025-03-31.
We continue collecting data during the review process to establish more
representative statistics over more diverse (Tor) network conditions. Thus, although we currently have a good baseline, it may be slightly modified in the final version.}.

\subsubsection{Component-level performance assessment}
\label{sec:prototype:component}

Our second approach focuses specifically on interactions within individual system elements that are critical for the user experience and thus bypasses preliminary setup processes like key exchange and initial sensor directory contact.
We conducted detailed timing analysis to identify performance bottlenecks in the prototype, aiming to minimize the interval between a user's arrival at a door and successful access grant (across 100 samples each):

\begin{itemize}
	\item Facial recognition processing time within the \textbf{sensor} (with a single face inside the frame) averaged 337\,ms ($\sigma$=42\,ms) on the laptop.
		We used a combination~\cite{bib:2023-hofer-momm} of ULFGFD\footnote{\url{https://github.com/Linzaer/Ultra-Light-Fast-Generic-Face-Detector-1MB}} and RetinaFace~\cite{bib:2020-retinaface} for face detection and ArcFace~\cite{bib:2019-arcface} for face recognition.
	\item \textbf{PIA} verification and transmission latency from sensor to verifier averaged 21.9\,ms ($\sigma$=2.61\,ms).
	\item \textbf{Verifier} validation and subsequent actions averaged 14.9\,ms ($\sigma$=1.16\,ms).
\end{itemize}

As the sensor accounts for over 90\% of the total processing time, we further decomposed the sensor processing pipeline to identify specific bottlenecks (again averaged over 100 samples):
85\,ms for the initial ULFGFD for generating face proposals, 166\,ms for RetinaFace for proposal confirmation and landmark detection, and 76\,ms for ArcFace for embedding creation.

Most time is spent after the initial scan of the image for faces, where each proposal must be confirmed and the embedding created.
While ULFGFD processing time remains constant regardless of scene complexity, RetinaFace and ArcFace components must be executed for each face proposal identified by ULFGFD.
This creates a potential scalability concern in scenarios with multiple subjects simultaneously present in the camera's field of view.

\subsubsection{Sensor scalability analysis}

Beyond basic performance metrics, we evaluate system scalability, particularly focusing on registration volume impacts and the number of faces within images as potential bottlenecks:

\paragraph{Registrations}
In considerations of global deployment, substantial user registrations may be associated with a single sensor.
Our implementation utilizes L2-norm calculation as the standard for comparing face embeddings.
Table~\ref{table:prototype:scale} shows the computational effect across registration volumes ranging from 1 to 10 million on a logarithmic scale
with a predictable linear scaling relationship. %
	
\begin{table}
\caption{Performance of matching an embedding to $N$ registrations and their storage requirement.}
\label{table:prototype:scale}
\begin{tabular}{rlll}
\toprule
Nr.\ of registrations & Mean {[}ms{]} & Std dev {[}ms{]}	& Storage \\
\midrule
1                    & \hphantom{00}0.000295      & 0.000026 & \hphantom{00}2 KiB\\
10                   & \hphantom{00}0.002935      & 0.000263 & \hphantom{0}20 KiB\\
100                  & \hphantom{00}0.029238      & 0.002632 & 200 KiB\\
1.000                & \hphantom{00}0.2962        & 0.022438 & \hphantom{00}2 MiB\\
10.000               & \hphantom{00}3.0235        & 0.050305 & \hphantom{0}20 MiB\\
100.000              & \hphantom{0}30.615        & 0.48046  & 200 MiB\\
1.000.000            & 303.06        & 3.5403   & \hphantom{00}2 GiB\\
\bottomrule
\end{tabular}
\end{table}
For context, when examining high-traffic areas, the largest airport by passenger count\footnote{\url{https://en.wikipedia.org/wiki/List_of_busiest_airports_by_passenger_traffic}}, the Atlanta International Airport, serves approximately 300,000 daily visitors (based on their reports\footnote{\url{https://www.atl.com/wp-content/uploads/2025/01/ATL-ATR-2412.pdf}} assuming a uniform distribution), while the highest frequented metro station, Tokyo's Shibuya Station, handles around 700,000 daily passengers\footnote{\url{https://www.tokyometro.jp/lang_en/corporate/enterprise/transportation/ranking/index.html}}.
These figures suggest one million registrations to represent a realistic upper bound for most real-world deployments.
Even accounting for potential false-positives (PIAs registering to sensors where users don't ultimately appear), it is unlikely that every passenger would register throughout an entire day at these high-traffic locations.
This scalability analysis confirms our implementation can handle registration volumes consistent with even the most demanding real-world deployment scenarios with acceptable performance characteristics.

\paragraph{Faces per image}

The computational complexity of our system scales with the number of faces detected in each frame, as both RetinaFace and ArcFace must process each face proposal individually.
From our component analysis in Section \ref{sec:prototype:component}, these two components require an average of 242\,ms per face.
Combined with the constant overhead of 85\,ms for generating proposals, we can derive the performance estimates given in Table~\ref{table:prototype:facescale}.
\begin{table}
	\caption{Processing time for varying numbers of faces detected in a single frame.}
	\label{table:prototype:facescale}
\begin{tabular}{rl}
\toprule
Nr.\ of faces visible & Mean {[}s{]} \\
\midrule
0                    & \hphantom{000}0.085      \\
1                    & \hphantom{000}0.327\\
10                   & \hphantom{000}2.505      \\
100                  & \hphantom{00}24.285      \\
1.000                & \hphantom{0}242 ($\approx$ \hphantom{0}4 min)\\
10.000               & 2420 ($\approx$ 40 min)\\
\bottomrule
\end{tabular}
\end{table}

These results reveal significant performance degradation as the number of faces increases, with processing times quickly becoming impractical for real-time applications.
Besides technical solutions to improve the performance, from an operational perspective, controlling the physical environment through measures such as gates or partitions can effectively limit the number of simultaneously visible faces to a manageable quantity.
This would ensure the system maintains acceptable response times while preserving core functionality in high-traffic scenarios.

\medskip
Summarizing, our initial scalability evaluations indicate linear scaling for registration matching with acceptable performance up to a few million users even on embedded hardware, sufficient for high-traffic venues if L2 is used as distance metric.
However, face detection processing time also increases linearly with the number of faces per frame, requiring environmental controls or technical optimizations in crowded settings to maintain real-time performance on embedded devices.

\subsection{Privacy Preserving Biometrics}
\label{sec:prototype:biometry}
All current implementations assume sufficient trust into the sensor to perform the final embeddings comparison and thresholding. 
However, even if the hardware RoT is intact (cf.\ section~\ref{sec:prototype:hw-rot}) and the software stack running on top is minimized, the whole chain is still complex and can in practice contain vulnerabilities. 
Therefore, another potential improvement for future work is not having to trust the sensor with any long-term stable biometric embeddings validated and integrated into credentials by issuing agencies: they are potentially problematic both as a perfectly linkable identifier for tracking (\ref{threat:data-collection}) and in aiding identity theft attacks (\ref{threat:identity-theft}).
Preliminary work towards implementing multi-party computation (MPC) of threshold matching based on the Funshade protocol~\cite{bib:2023-funshade} has recently been completed~\cite{bib:2024-bodi-masterthesis}.
This allows both the PIA and sensor to keep their respective embedding input for the matching computation confidential and therefore further reduces required trust into the sensor.
Under the assumption of previous parameter setup and with current optimizations, a single MPC is already fast enough to be insignificant from a performance point of view ($120 \mu s$ local computation latency on a laptop~\cite{bib:2024-bodi-masterthesis}), but evaluating scalability in the range of PIA-sensor registrations across the network and MPC interactions is still subject to future work.

\section{Formal Verification}
\label{sec:formalverification}
We now introduce our formal model of Digidow including relevant preliminaries for the verification approach we use.
Our goal with this formal model is to create an abstract representation of our protocol design to verify the expected behavior. 
Thus, we abstract away some technical implementation details, such as cryptographic signatures or remote attestation to create a cleaner formal model. 
This abstraction includes privacy properties, which will be addressed in future work. 
Our formal model can be found here \attachfile{digidow-formal-verification/digidow.spthy}.

\subsection{\textsc{Tamarin} Prover}
We use \textsc{Tamarin}, a security protocol verification tool, to formally model and verify the Digidow architecture.
Our formal model consists of several state transitions representing the protocol flow of the proposed architecture. 
These states can be modeled in \textsc{Tamarin} by using so-called \emph{facts} in the form of \(F(t_1..t_n)\), where \(F\) is the name of the fact and \(t_i\) represents data transmitted from one state to another. 
\textsc{Tamarin} also provides some built-in facts, such as \(In(..)\) and \(Out(..)\) that can be used to represent the adversary network. 
We use the Dolev-Yao adversary model, where we do not limit the capabilities of an adversary with regard to eavesdropping, tampering, or replaying messages within this network. 
The actual state transitions are modeled by using multiset rewrite rules (MSR). 
A MSR consists of four relevant parts: a name, a left-hand side part and an \emph{action fact} and a right-hand side part. 
Each MSR is applied as soon as the state of the left-hand side is satisfied and transitions this state to the right-hand side state. 
The \emph{action fact} (the middle of the state transition) is used to label the transition so that we can reference it later during verification of our security properties. 
Finally, the most important task is to verify relevant security properties.
These properties can be described in \textsc{Tamarin} by using \emph{lemmas} that consist of a name and a first-order formula. 
The \emph{lemmas} in \textsc{Tamarin} can either be defined in a way to verify if there exists a trace or whether all traces fulfill certain criteria. 
The criteria used in the verification stage are based on the previously defined \emph{action facts} and some additional verification step of the respective data.

\subsection{Trust Assumptions}
We adhere to the same trust assumptions outlined in section~\ref{sec:threatmodel:trustassumptions} for our formal model, and particularly that the secure hardware works as specified and is not broken (\ref{trust:secure-hardware}). 
Thus, we assume that sensor detection is always trustworthy, as attestation is done before using the sensor.

\subsection{Security Properties}
Our initial formal model is primarily based on two specific threats: \ref{threat:identity-theft} (Identity theft) and \ref{threat:identity-modification} (Identity modification).
Each security property listed below is verified within this formal model and consists of two proofs: one demonstrating that the specific security property can be compromised in case our proposed mitigation controls are not utilized, and another to ensure that \textsc{Tamarin} does not find any trace where an adversary would be successful when using the proposed architecture. 

\begin{enumerate}[start=1,label={L\arabic*}]
\item \textbf{Identity Spoofing (\ref{threat:identity-theft}):}
\label{sec:formalverification:identity-spoofing}
The identity of an individual is one of the most important security properties to be protected. 
Thus, we consider the attack scenario where an adversary tries to issue certain attributes for another individual. 
The trust assumption for this particular case is that the issuing authority verifies the identity before providing requested attributes (\ref{trust:stakeholder-interest}.\ref{trust:stakeholder-interest:issuer-verifer} and \ref{trust:stakeholder-interest}.\ref{trust:stakeholder-interest:issuer-owner}). 
Our formal model includes two proofs for this security property: (1) verifies whether an adversary can successfully request attributes if verification is not performed, while (2) assumes verification is done and ensures the adversary cannot issue such attributes.
Specifically for (1) we prove that \(\exists\ A,i,a,b,ba : \neg(ba = b)\) and for (2) we verify that \(\forall\ A,i,a,b,ba : (b = ba) \), where \(A\) is the individual, \(i\) the identity, \(a\) the attributes, \(b\) the biometry, and \(ba\) the biometry issued by the adversary. 

\item \textbf{Credentials Spoofing (\ref{threat:identity-modification}):} 
\label{sec:formalverification:credential-spoofing}
An individual's biometry is an essential and unique property to actually authenticate them and is associated with the respective PIA. 
We intentionally allow everyone to create a PIA including biometric information, including adversaries. 
However, our protocol ensures that a PIA created or tampered by an adversary cannot be linked to attributes issued to another individual, preventing the adversary, e.g., from abusing assigned privileges.
To prove that this type of attack is not possible in Digidow, we create two proofs: (1) verifies whether an adversary can successfully assign arbitrary attributes to the own PIA without verification of the assigned biometry, while (2) verifies the biometry (assuming \ref{trust:biometry}) so that the spoofing attempt is detected.
Specifically for (1) we prove that \(\exists\ A,S,i,p,a,aa,b,ba,s : (\neg(b = ba) \wedge (a = aa))\) and for (2) we verify that \(\forall\ A,S,i,p,a,aa,b,ba,s : (b = ba)\), where \(A\) is the individual, \(S\) the sensor, \(i\) the identity, \(p\) representing the PIA, \(a\) the attributes, \(aa\) the attributes provided or manipulated by the adversary, \(b\) the biometry, \(ba\) the biometry issued by the adversary, and \(s\) the sensor id. 

\item \textbf{Sensor Detection (\ref{threat:identity-theft},\ref{threat:identity-modification}):}
\label{sec:formalverification:sensor-detection}
The third security property we consider within our initial model is the trustworthiness and reliance with regard to identification of individuals. 
We assume sensors to be trustworthy as they can be attested remotely for verification purposes (\ref{trust:secure-hardware}). 
However, we model the scenario where an adversary tries to spoof or tamper an identity in front of a sensor, e.g., by using a mask or just a picture of somebody else. 
We verify our protocol design with two additional proofs: (1) verifies whether an adversary can successfully spoof an identity in front of a sensor without depending on detection accuracy, while (2) assumes accurate detection and ensures that the adversary cannot spoof the target individual's identity (\ref{trust:biometry}).
Specifically for (1) we prove that \(\exists\ A,S,i,p,b,bd,br,s,sd,sr : (\neg(bd = br))\) and for (2) we verify that \(\forall\ A,S,i,p,b,bd,br,s,sd,sr : (bd = br) \), where \(A\) is the individual, \(S\) the sensor, \(i\) the identity, \(p\) representing the PIA, \(b\) the biometry, \(bd\) the biometry detected by a sensor, \(br\) the biometry registered with a PIA, \(s\) the sensor id, \(sd\) the sensor that detects a biometry, and \(sr\) the sensor that registers a PIA. 
\end{enumerate}

\section{Comparison}
\label{sec:comparison}
We now compare our proposed architecture with related deployed systems or architecture references currently being discussed for standardization.

\subsection{Aadhaar}
Aadhaar~\cite{bib:aadhaar} represents the world's largest digital identity system operated by the \emph{Unique Identification Authority of India} (UIDAI). 
It was designed to empower Indian citizens with a unique identity that can be used anywhere at any time. 
These identities are stored in a central identities data repository (CIDR) operated by UIDAI, who in turn grants relevant stakeholders access to this database if necessary. 
Aadhaar number holders (identity owners) can use the information stored about them within the CIDR to authenticate themselves. 
Registrars (identity issuers) can enroll new individuals into the CIDR database and provide them with valid attributes. 
Authentication user agencies (identity verifiers) can verify authentication data by comparing it to the data stored within the CIDR. 
Authentication devices (biometric sensors) are operated by authentication user agencies or registrars and have to follow guidelines specified by the UIDAI. 

There are several aspects of Aadhaar that are especially relevant when comparing it to other digital identity systems:
First, it does not support biometric identification.
Every authentication request sent to the CIDR must contain the 12-digit Aadhaar number uniquely identifying a single individual~\cite{bib:aadhaar-operation-model}. 
In practice, this means that using Aadhaar requires an Aadhaar card that contains the Aadhaar number in machine-readable format, so authentication devices can automatically read it. 
Second, the CIDR only responds to requests with Yes or No, it does not share attributes (mitigating \ref{threat:data-oversharing}), which forces authentication user agencies to send very specific requests\footnote{There is one exception to this: via the use of the electronic know-your-customer (eKYC) functionality.} (e.g., ``does the person with number X have the following fingerprint, and if yes, is that person eligible for social benefits?'') ~\cite{bib:aadhaar-operation-model}.
Third, communication with the CIDR is limited to trusted agencies~\cite{bib:aadhaar-asa}. 
The UIDAI certifies authentication service agencies which are considered trusted with securing network traffic before forwarding it to the CIDR. 
Authentication user agencies have to send all of their requests via such an authentication service (or become one themselves). 

In comparison to the Digidow architecture, Aadhaar requires its users to fully trust the UIDAI.  
The UIDAI can not only directly modify any information stored in the CIDR (realizing \ref{threat:identity-theft}), it is also responsible for appointing registrars, authentication user agencies, and authentication service agencies. 
If any of them are compromised in the future, the UIDAI is responsible for detecting malicious behavior and for compensating for any damage done by them. 
Reports~\cite{bib:2018:khaira:aadhaar-breach,bib:2019:vudali:aadhaar-details-found} about several data breaches linked to Aadhaar---that are most likely tied to either insufficient security measures or corruption---indicate that this is not only a theoretical assumption. 

Even if the data within the database is kept private, monitoring accesses to the CIDR enables the UIDAI to track when and how Indian citizens utilize their identities (realizing \ref{threat:data-collection}). 
This possibility is acknowledged and regulated in clause 33 of the Aadhaar Act~\cite{bib:2016:aadhaar-act,bib:2019:aadhaar-act-amended} which explicitly allows the disclosure of identity information and authentication records in the interest of national security. 

As the single controlling instance, the UIDAI, could also exclude individuals, companies or even entire ethnic groups from obtaining or using their Aadhaar identities (realizing \ref{threat:denial-of-service}). 
Due to the structure of requests made to the CIDR, they could even prevent individuals from accessing specific services without removing them from the system entirely.
The Aadhaar Act explicitly states that every Indian citizen is entitled to an Aadhaar number, and---so far---there have been no confirmed instances of citizens selectively being denied enrollment into Aadhaar. 
However, clause 7 of the Aadhaar act also mandates that citizens not yet enrolled in Aadhaar must not be denied access to government services and there are numerous reports about this happening~\cite{bib:2014:sood:aadhaar-no-lpg-subsidy-without,bib:2017:porecha:aadhaar-needed-for-abortion,bib:2022:chandran:aadhaar-needed-for-school}. 
Those complaints are so common that the UIDAI has repeatedly been forced to make public statements~\cite{bib:2018:uidai:no-denial} reminding other government agencies that the lack of an Aadhaar number does not justify excluding genuine beneficiaries from essential services and threatening legal actions against continued violations. 
From a design perspective, this highlights a general issue with legal protections: 
They are only useful if they can be enforced---and the groups and individuals most likely to be discriminated against often lack the financial and legal resources required to enforce their rights through a judicial process. 

Despite those limitations, Aadhaar is a very successful digital identity system, that---after initially being introduced as optional---is now mandatory for all citizens and continuously expands into new areas.
Allowing banks and telecommunications providers to authenticate their users via Aadhaar~\cite{bib:2019:thaker:aadhaar-private-access} and using Aadhaar to access medical and health information~\cite{bib:2018:tibbitts:aadhaar-health-data} are just two examples for this trend. 
The potential for quick and massive expansion demonstrated by Aadhaar confirms one of Digidow's core assumptions: 
Digital identity systems need strong security, safety and privacy guarantees built into them on a technical level from the beginning because successful systems will be continuously expanded.

\subsection{Mobile Driving License}
\label{sec:comparison:mdl}
Mobile driving licenses (mDL) are an approach to bring well-standardized physical driving license documents (cf.\ international standards series ISO/IEC 18013) to mobile devices.
ISO/IEC 18013-5~\cite{bib:2021:iso:18013-5} defines data structures and communication protocols that allow an individual (identity owner) to present a driving license (identity attributes) carried on a mobile device to arbitrary verifiers in the physical world; covering use-cases in the public and private sectors.
Whereas the initial standardization project had its focus primarily on presenting driving privileges, the communication protocols specified in ISO/IEC 18013-5 were already designed with more general identity documents in mind and even permit the addition of arbitrary (identity) attributes~\cite{bib:2021:iso:18013-5}.
This is due to the fact that, in many countries, driving licenses are often also used as a means of identification (even in jurisdictions where they are not considered to be identity documents).

Similar to Digidow, an ISO mDL enables physical-world verification use-cases with a digital identity.
The standard implements several privacy mechanisms that Digidow also expects from a digital identity.
Among them are the selective disclosure of attributes (mitigates \ref{threat:data-oversharing}) and unlinkability (mitigates \ref{threat:data-collection}) across all protocol layers.
With regard to selective disclosure, ISO mDL also introduces a mechanism for age attestation without revealing the actual date of birth (through a series of ``age over X'' queryable attributes).
Besides randomization of addresses in underlying transport protocols, ISO mDL also addresses unlinkability of an identity document statically signed as a whole (mitigates \ref{threat:identity-modification}) by its issuer:
\begin{enumerate*}
\item it introduces randomizers to diminish linkability and guessability of non-disclosed attributes through static signatures that are required to be transmitted for verification;
\item it permits issuance of single-use mobile security objects (MSO; static signature blob over an mDL) to prevent linkability over multiple presentations of the same mDL.
\end{enumerate*}
Besides offline-verifiable mobile driving licenses (where the identity document is stored on the mobile device), ISO mDL also supports an online scenario, where the identity verifier obtains authenticated identity attributes directly from an issuing authority upon verification (the mobile device acts only as transaction enabler).
The online scenario enables the issuing authority to track usage of each mDL (realizing~\ref{threat:data-collection}).

The standardization efforts started around mobile driving licenses evolved into a broader standardization effort around building blocks for identity management via mobile devices (cf.\ international standards and technical specifications series ISO/IEC 23220\footnote{Further technical specifications in ISO/IEC 23220 are still under detailed discussions at the time of this writing. Whereas, e.g., the concept of single-use MSOs is made possible in published standards, upcoming texts are expected to strongly recommend their use.}).
These efforts resulted into publication of ISO/IEC 23220-1~\cite{bib:2023:iso:23220-1}, which defines design principals and system architectures for mobile electronic identity documents.
The standard defines both on-site (physical world) identification and remote (digital world) identification uses-cases for a mobile eID.
Its privacy and security principles closely follow those of mDL.
With regard to privacy, data minimization (through partial attribute release, unlinkability, and domain-specific pseudonyms) and user-centricity (through informed consent and choice) are core design principles that we also pick up in the Digidow architecture.
With regard to security, the standard particularly considers accuracy and quality of digital identities (through user binding, authenticity and integrity mechanisms, and update and revocation mechanisms), and information security in general (through secure data storage, eavesdropping, and cloning protection).

In particular the approach for binding an individual to their digital identity significantly differs from the approach envisioned in the Digidow architecture:
ISO/IEC 23220-1 expects the \emph{verifier} to verify the binding between a digital identity document and its holder.
A typical approach is to transmit (biometric) attributes that permit a verifier to establish such a link; e.g.\ ISO/IEC 18013-5 requires the facial image to be always transmitted to a verifier for visual face matching.
Sharing such attributes, in turn, introduces linkability through matching such attributes (mitigating \ref{threat:identity-theft} but enabling~\ref{threat:data-collection} if such information is abused by the verifier).
Digidow tries to eliminate this privacy impact by outsourcing the biometric matching to a sensor infrastructure that PIAs sufficiently trust for exchanging biometric information and verifiers sufficiently trust for establishing the link between the detection of a person in the physical world and its corresponding digital identity (cf.\ section~\ref{sec:challenges:unlinkability:trusting-sensors}).

\subsection{Digital Travel Credentials}
ICAO digital travel credentials (DTC)~\cite{bib:2020:dtc-guidelines-4.4} decouple the data usually stored on the chip of an electronic machine-readable travel document (eMRTD) from the actual travel document (passport booklet).
The travel document is split into a virtual component (DTC-VC)~\cite{bib:2020:dtc-vc-1.2} that encapsulates signed identity attributes and a physical component (DTC-PC) that travelers (identity owners) may use as an additional authenticator to prove their authorization to use a certain DTC-VC, opening travel documents to new form factors besides passport booklets.

DTC aim to make air travel more efficient by allowing travelers (identity owners) to share their DTC-VC (identity attributes) ahead of travel with verifiers (among them are airlines and border control authorities).
Verifiers can use these pre-shared attributes to prepare for the actual travel and border-crossing in order to guarantee a seamless passenger flow.
One particular preparatory task is to bootstrap biometric attributes for later identification.
Unlike with Digidow, the whole set of identity attributes is transmitted directly to the verifier and DTC-VC are neither designed for partial disclosure of subsets of attributes nor to eliminate linkability between presentations of a credential (mitigating \ref{threat:identity-modification} but realizing \ref{threat:data-collection} and \ref{threat:data-oversharing}).
This also includes attributes with biometric features (mitigating \ref{threat:identity-theft}).
Although the guidelines for DTCs~\cite{bib:2020:dtc-guidelines-4.4} suggest that only facial biometrics should be included in a DTC-VC, the technical report on DTC-VC~\cite{bib:2020:dtc-vc-1.2} declares other biometric data groups (e.g.\ for fingerprints) as optional fields.
This means that verifiers will receive (and are capable of storing) these highly sensitive biometric attributes.

\subsection{eIDAS}
The European Union has recognized the importance of digital identities for providing digital goods and services. 
Since there were already several different digital identity systems deployed on the national level, the eIDAS regulation~\cite{bib:2014:EU-eIDAS} opted not to introduce a new pan-European system but aims for compatibility between existing national identity systems.%
Therefore, eIDAS---much like the proposed Digidow architecture---can be considered a distributed digital identity system. 
Member states act as identity issuers for their citizens, either by directly issuing identities or by delegating this responsibility to a trusted party. 
If member states act as identity verifiers for providing public services, %
they have to accept every eIDAS identity, independent of the issuing country. 

Trust into identity issuers is ensured by the European Commission. 
New identity systems as well as changes to existing ones are notified to the Commission, which verifies their adherence to the eIDAS regulation as well as minimal technical requirements needed to meet certain assurance levels. 
Once an identity issuer is confirmed as trustworthy by the Commission, it must be accepted by public identity verifiers operated by all member states.
Identity verifiers not operated by the public sector %
are not mandated to accept digital identities from all issuing authorities. 
Instead, they have the freedom to decide which identity issuers they want to trust. 
This approach of granting freedom of choice to verifiers can also be found in the proposed Digidow architecture. 

Nevertheless, there exist significant differences between eIDAS and Digidow. 
Most importantly, eIDAS does not address the challenge of using digital identities in the physical world. 
Its primary goal is to strengthen the digital market; and a purely digital market has little need for physical world interactions. 
This also explains why there is no notion of sensors or biometry in this section. 

The second important difference is that eIDAS is built on top of already existing digital identity systems. 
Apart from minimum technical requirements for assurance levels, member states are free to design their identity systems however they please. 
Aadhaar and the Digidow architecture would be two potential starting points for such designs. 
Despite their differences, they all enable the same access to public (and hopefully also private) services. 

The latest iteration of eIDAS proposes the introduction of user-controlled digital wallets to store identity information~\cite{bib:2024:EU-eIDAS2}. 
To remain consistent with established European data privacy regulations, wallets are required to ``enable privacy preserving techniques which ensure unlinkability, where attestation of attributes do not require the identification of the user''~\cite{bib:2024:EU-eIDAS2} (mitigating~\ref{threat:data-oversharing}). 
This aligns perfectly with the discussion presented in section~\ref{sec:challenges:unlinkability}. 
Should eIDAS be extended to support interactions in the physical world, these unlinkability requirements would be very hard to fulfill with current centralized systems like Aadhaar. 

\subsection{Summary}
\begin{table}
	\caption{Comparison of threats addressed in existing identity systems and Digidow with linkage to mitigations and formally verified lemmas}
	\label{table:comparison}
    \newcommand{\mitigatedTechnical}{\ding{51}}%
    \newcommand{\mitigatedLegal}{\ding{108}}%
    \newcommand{\unmitigated}{\ding{55}}%
    \newcommand{\unknown}{\bfseries ?}%
    \newcommand{\futureWork}{\ding{115}}%
	\begin{tabular*}{\textwidth}{>{\raggedright\arraybackslash}p{25mm}>{\centering\arraybackslash}p{19mm}>{\centering\arraybackslash}p{19mm}>{\centering\arraybackslash}p{19mm}>{\centering\arraybackslash}p{19mm}>{\centering\arraybackslash}p{16mm}}
		\toprule
		& \ref{threat:identity-theft} & \ref{threat:identity-modification} & \ref{threat:data-collection} & \ref{threat:data-oversharing} & \ref{threat:denial-of-service}\\
		& \small (identity theft) & \small (tampering) & \small (data coll.) & \small (oversharing) & \small (denial)\\
		\midrule
		Aadhaar         & \mitigatedTechnical$^1$ & \mitigatedTechnical & \unmitigated & \mitigatedTechnical & \mitigatedLegal \\
		ISO/IEC 18013-5 & \mitigatedTechnical & \mitigatedTechnical & \mitigatedLegal & \mitigatedTechnical & \mitigatedTechnical$^2$ \\
		ICAO DTC        & \mitigatedTechnical & \mitigatedTechnical & \unmitigated & \unmitigated & \unmitigated$^3$ \\
		eIDAS (2024)    & \unknown & \mitigatedTechnical & \unknown & \mitigatedTechnical & \mitigatedLegal \\
		\midrule
		Digidow         & \mitigatedTechnical & \mitigatedTechnical & \mitigatedTechnical & \mitigatedTechnical & \mitigatedTechnical \\
		with mitigations in~sections: & \ref{sec:protocol:issuing-attributes}, \ref{sec:protocol:transaction}, \ref{sec:prototype:hw-rot}, \ref{sec:prototype:cryptography}, \ref{sec:prototype:biometry}
                        & \ref{sec:protocol:issuing-attributes:initial-attributes}, \ref{sec:protocol:transaction}, \ref{sec:prototype:cryptography}
                        & \ref{sec:protocol:sensor-directory}, \ref{sec:protocol:transaction}, \ref{sec:prototype:network-privacy}, \ref{sec:prototype:hw-rot}, \ref{sec:prototype:cryptography}, \ref{sec:prototype:biometry}
                        & \ref{sec:protocol:transaction}, \ref{sec:prototype:cryptography}
                        & \ref{sec:protocol:issuing-attributes}, \ref{sec:protocol:transaction}, \ref{sec:prototype:network-privacy}, \ref{sec:prototype:hw-rot} \\
    formally verified:    & \ref{sec:formalverification:identity-spoofing}, \ref{sec:formalverification:sensor-detection} & \ref{sec:formalverification:credential-spoofing}, \ref{sec:formalverification:sensor-detection} & \futureWork & \futureWork & - \\
		\bottomrule
	\end{tabular*}

    \footnotesize
    \smallskip
	\begin{tabular*}{\textwidth}{>{\centering\arraybackslash}p{1.5em}@{}>{\raggedright\arraybackslash}p{\dimexpr\textwidth-1.5em\relax}}
        \mitigatedTechnical     & Mitigation through technical measures\\
        \mitigatedTechnical$^1$ & Mitigation, but huge attack surface\\
        \mitigatedTechnical$^2$ & Limited scope through offline scenario and expected fallback to physical ID in current implementations\\
        \mitigatedLegal         & Mitigation through legal/organizational measures\\
        \unmitigated            & No mitigation\\
        \unmitigated$^3$        & Several countries are known to refuse issuing travel documents\\
        \futureWork             & Future work requiring further development of formal verification for privacy properties \\
        \unknown                & Not yet defined\\
        & \\
	\end{tabular*}
\end{table}
Table~\ref{table:comparison} provides an overview over the digital identity systems presented in this paper and the threats they address including the linkage between formally verified lemmas and the corresponding threats. 
The original version of eIDAS was excluded because it is not a digital identity system itself and could not be evaluated w.r.t.\ the threats listed in section~\ref{sec:threatmodel:threats}.
The current draft for a new version of eIDAS, while still incomplete, already contains enough concepts to at least partially be included the table. 
As this analysis is based on incomplete current drafts, it does not provide a foundation for further analysis, but it provides an idea of the direction the European Union intends to take regarding digital identity systems. 

When it comes to how threats are being mitigated, we distinguish between technical protections that make it infeasible to realize a threat, and legal measures that forbid or restrict who can realize a threat. 
We consider legal protections weaker than technical ones, because they are easier to change retroactively\footnote{A single political election, coup, or abuse of existing legal powers can immediately invalidate all such legal mitigations.} and require users to trust in their identity system providers.

\section{Conclusion}
\label{sec:conclusion}
In this article, we proposed an architecture for creating a distributed digital identity system for authenticating and authorizing actions of individual persons in the physical world;
the most prominent use cases include opening doors and other physical barriers to allow individuals access to services they are authorized to use, without forcing them to carry any form of identity document or mobile device with them.
Through the implied reliance on biometric authentication, an interesting intersection of challenges arises between anti-forgery, network privacy, biometric privacy, cryptographic privacy, trusting ubiquitous sensors embedded into the environment, and scalable network service discovery in a globally distributed system.
In addition to the better-known security requirements for digital authentication and authorization systems, \emph{unlinkability} emerges as a core privacy requirement underlying many of the more specific challenges.

We propose an abstract distributed protocol between three parties---(biometric) sensors, verifiers, and a Personal Identity Agent (PIA) as digital representative of each individual---to address these challenges on an architectural level, most importantly by avoiding centralized points of failure, surveillance, and control.
Recent developments like the upcoming eIDAS 2.0 regulation confirm the real-world demand for distributed digital identity systems with strong privacy protections. 
However, some challenges cannot be addressed through the protocol architecture alone, but depend on the specific implementation.
Therefore, we also present a feasibility evaluation with partial prototypes and a discussion of implementation options where external dependencies such as international identity format standards are still under development.

Particularly, 
\begin{enumerate}[label=(\alph*)]
	\item the selection of an international standard for cryptographic signatures over identity attributes that provides unlinkability even against collusion between issuers and verifiers,
	\item in-production use of remote attestation of boot-time and run-time state of embedded sensor devices with hardware roots of trust, 
	\item biometric comparisons between two parties revealing only if the calculated distance is below a certain threshold---especially wrt.\ scalability concerns, 
	\item location prediction models capable of reliably predicting individuals next location based on their previous activities,
	\item further improving end-to-end latency and scalability of private network communication such as through Tor onion service, and
	\item finalizing global deployment standards in terms of interoperability and data governance
\end{enumerate}
are subject to future work before a global deployment of digital identity systems for physical services with our described high level of privacy and security guarantees becomes practical.

\section*{Declarations}

\paragraph{Availability of data and materials.}
Data collected for the purposes of this paper (as used in section~\ref{sec:prototype:implementation}) is limited to timing measurements that are listed within this paper, but source code of all prototype parts mentioned in \autoref{sec:prototype} are available under open source licenses.

\paragraph{Authors' contributions.}
René Mayrhofer developed the original vision, architecture, and wrote the initial project proposal as well as main parts of this article. Michael Roland was instrumental in all parts of research leading to the current status and in managing the research group, with particular contributions to the overall architecture and threat model. Tobias Höller contributed especially to network protocol definitions, definitions of unlinkability, and the comparison to related work. Philipp Hofer implemented the sensor and face recognition pipeline and designed and executed the analysis of the performance measurement experiments for characterizing the proof-of-concept prototype. Mario Lins contributed mainly to formulating the structured threat model and formal verification. All authors contributed to the overall presentation and text in this article.

\begin{acks}
This work has been carried out within the scope of Digidow, the Christian Doppler Laboratory for Private Digital Authentication in the Physical World. 
We gratefully acknowledge financial support by the Austrian Federal Ministry of Labour and Economy, the National Foundation for Research, Technology and Development, the Christian Doppler Research Association, 3 Banken IT GmbH, ekey biometric systems GmbH, Kepler Universitätsklinikum GmbH, NXP Semiconductors Austria GmbH \& Co KG, and Österreichische Staatsdruckerei GmbH.

We would like to thank Stefan Rass, Daniel Slamanig, and Michael Sonntag for detailed feedback on earlier drafts of this article, and particularly Gerald Schoiber, Martin Schwaighofer, and Stefan Kempinger for their in-depth contributions to the overall project and many of the implementations that form part of the prototype. %
\end{acks}

\bibliographystyle{ACM-Reference-Format}
\bibliography{literature}

\end{document}